\documentclass[draft]{agujournal2019}
\usepackage{url} 
\usepackage[finalnew]{trackchanges} 
\usepackage{soul}
\usepackage{setspace}
\usepackage{amsmath,amssymb}
\usepackage{bm}
\usepackage{color}
\usepackage{here}
\usepackage{multirow}

\addeditor{1stRev}

\draftfalse

\journalname{Journal of Advances in Modeling Earth Systems (JAMES)}

\begin{document}

%
%


\title{Spatio-Temporal Super-Resolution Data Assimilation (SRDA) Utilizing Deep Neural Networks with Domain Generalization}

%
%




\authors{Yuki Yasuda\affil{1} and Ryo Onishi\affil{1}}


\affiliation{1}{Global Scientific Information and Computing Center, Tokyo Institute of Technology, 2-12-1 Ookayama, Meguro-ku, Tokyo 1528550, Japan}




\correspondingauthor{Yuki Yasuda}{yasuda.y.aa@m.titech.ac.jp}




\begin{keypoints}
\item We propose a neural network-based scheme for data assimilation and spatio-temporal super-resolution using physics-based simulations
\item A data-augmentation technique was developed to improve the robustness of inference in the fusion of data-driven and physics-based models
\item A test with idealized ocean jets showed that the proposed approach efficiently infers high-resolution results with ensuring the robustness
\end{keypoints}

%
%

%
%


\begin{abstract}
Deep learning has recently gained attention in the atmospheric and oceanic sciences for its potential to improve the accuracy of numerical simulations or to reduce computational costs. Super-resolution is one such technique for high-resolution inference from low-resolution data. This paper proposes a new scheme, called four-dimensional super-resolution data assimilation (4D-SRDA). This framework calculates the time evolution of a system from low-resolution simulations using a physics-based model, while a trained neural network simultaneously performs data assimilation and spatio-temporal super-resolution. The use of low-resolution simulations without ensemble members reduces the computational cost of obtaining inferences at high spatio-temporal resolution. In 4D-SRDA, physics-based simulations and neural-network inferences are performed alternately, possibly causing a domain shift, i.e., a statistical difference between the training and test data, especially in offline training. Domain shifts can reduce the accuracy of inference. To mitigate this risk, we developed super-resolution mixup (SR-mixup)--a data augmentation method for domain generalization. SR-mixup creates a linear combination of randomly sampled inputs, resulting in synthetic data with a different distribution from the original data. The proposed methods were validated using an idealized barotropic ocean jet with supervised learning. The results suggest that the combination of 4D-SRDA and SR-mixup is effective for robust inference cycles. This study highlights the potential of super-resolution and domain-generalization techniques, in the field of data assimilation, especially for the integration of physics-based and data-driven models.
\end{abstract}

\section*{Plain Language Summary}

One challenge in the Earth sciences is the simulation of atmospheric and oceanic processes with high accuracy or at high spatio-temporal resolution. The neural network, a model developed in the field of artificial intelligence, has recently gained attention for its potential to learn the relationship between any variables from a sufficient amount of data. For example, a neural network can be trained to make high-resolution images from low-resolution images. This process is called super-resolution. We propose the incorporation of a neural network for super-resolution into an atmospheric or ocean model that performs numerical simulations at low resolution. Since the atmospheric or ocean model is of low resolution, the proposed framework is computationally efficient while still allowing for high-resolution results with the aid of the neural network. Furthermore, to make the inference accurate, observation data are fused with atmospheric or ocean predictions in the neural network. In real-world applications, it is important to make predictions robust to noise from various factors. Accordingly, we developed a new technique where certain noise is added during the training of the neural network. The approach proposed here can efficiently compute high-resolution predictions of the atmosphere and ocean without compromising robustness.

%
%

%


%
%
%
%

\section{Introduction}  \label{sec:introduction} 

Numerical simulation is an essential tool in the atmospheric and oceanic sciences. Despite the ever-increasing computer resources, it may still be difficult to conduct numerical simulations at high resolution or with high accuracy. One solution is to employ data-driven models such as neural networks (NNs) \cite{Brunton+2020ARFM, Duraisamy2021PRF, Kashinath+2021PTRS}. For instance, NNs have been used for surrogate modeling of the Navier-Stokes equations, subgrid-scale modeling of Reynolds stress tensors, and super-resolution of fluid flows. These NNs can make numerical simulations more efficient or more accurate.

Super-resolution (SR) refers to techniques that infer high-resolution (HR) images from low-resolution (LR) images. Methods based on deep learning have been studied in computer vision \cite<e.g.,>{Dong+2014ECCV, Ha+2019, Anwar+2020ACMCS}. The success of such NNs has promoted the applications of SR to atmospheric and oceanic data \cite<e.g.,>[]{Ducournau+2016PRRS, Vandal+2017ACM, Onishi+2019SOLA, Stengel+2020PNAS, Wang+2021GMD, Wu+2021GRL, Yasuda+2023BAE}. One example is SR simulation \cite{Onishi+2019SOLA, Wang+2021GMD, Wu+2021GRL}. With this scheme, time evolution is calculated from LR simulations using a physics-based model, and HR inferences are obtained by inputting the LR results into a trained NN. Super-resolution simulation is computationally efficient because it requires only LR simulations using a physics-based model, and the elapsed time of inference by a NN is usually negligible. However, there is a concern that LR simulations may deviate from the true state because of unresolved processes. Data assimilation may be useful for reducing such deviation.

Data assimilation (DA) is a framework that estimates the state of a system based on observations and background states provided by a model \cite{Asch+2016SIAM}. Methods of DA are typically variational or statistical, with NNs being applied to both types. Recent progress in the application of NNs to DA has been summarized in a review \cite{Cheng+2023arXiv}. Variational methods, such as 4D-Var, are analogous to deep learning, in that both employ adjoint backpropagation \cite{Abarbanel+2018NC, Chen+2018NIPS}. The minimization of variational cost can be conducted with NNs, achieving more precise 4D-Var estimations \cite{Fablet+2021, Beauchamp+2022GMDD}. Neural networks have also been incorporated into statistical methods, such as ensemble Kalman filters (EnKFs). To reduce the computational cost of calculating ensemble evolution, the system model can be replaced with a NN \cite{Ouala+2018RS, Chattopadhyay+2022GMDD}. In addition, the analysis step can be conducted in low-dimensional latent spaces to further increase computational efficiency \cite{Amendola+2021, Peyron+2021QJRMS, Liu+2022EABE}. Instead of replacing system models, ensemble members can be obtained using generative models, such as variational autoencoders (VAEs), which also reduce computation time \cite{Groom2021QJRMS, Yang+2021JCP}.

Super-resolution has recently been included in DA processes \cite{Barthelemy+2022OD}, as super-resolution data assimilation (SRDA), which applies a statistical DA method, such as EnKF, to flow fields super-resolved by a NN. SRDA can be interpreted as a fusion of SR simulation and DA. Specifically, the time evolution of an ensemble is obtained from LR simulations via a physics-based model, as in SR simulation. The NN independently super-resolves each LR result in the ensemble. Observations are then assimilated into these super-resolved HR data. The obtained HR analysis is mapped to the LR space using algebraic methods, such as linear interpolation, and this mapped state provides the initial condition for the next assimilation cycle. SRDA performs super-resolution and data assimilation separately, making it easy to incorporate into an existing DA framework. Some studies have shown that NNs can perform the entire or a part of the analysis step. For instance, NNs can directly infer the analysis states \cite{Krishnan+2015arXiv, Fraccaro+2017NIPS} or can estimate background error covariance matrices \cite{Ouala+2018RS, Groom2021QJRMS}. These studies imply the potential for simultaneous SR and DA. However, to the authors' knowledge, SR has yet to be fully studied in the context of DA and such a proof of concept has not been demonstrated.

In SRDA \cite{Barthelemy+2022OD}, physics-based simulations and NN inferences are alternated, potentially causing a domain shift \cite<e.g.,>{MorenoTorres+2012PR}, in which the statistical distribution differs between the training and test data. Domain shifts can occur whenever physics-based simulations and NN inferences are interdependent. For instance, if training data are generated in advance via physics-based simulations (i.e., offline training), a domain shift will occur because the training data are independent of the NN inference. Although this issue is briefly mentioned by \citeA{Barthelemy+2022OD}, it has not been investigated. Online training provides a potential solution to this domain shift \cite{Rasp2020GMD}. In this approach, training data are updated during the optimization of NNs through physics-based simulations \cite<e.g.,>{Krasnopolsky+2020arXiv}. However, online training may be computationally intensive and is not always feasible.

Domain generalization aims to improve generalizability across datasets with different distributions and to overcome domain shift problems \cite<e.g.,>{Wang+2022IEEE, Zhou+2022IEEE}. Domain generalization is similar to transfer learning, but differs in access to the test-data distribution. In transfer learning, a pre-trained NN is fine-tuned using small samples with the same distribution as the test data. In contrast, in scenarios of domain generalization, fine-tuning is impossible because such samples are not available. In other words, the test-data distribution remains unknown up to the test phase. Domain-generalization methods may be useful for fusing data-driven and physics-based models; however, the applicability of domain generalization in applying NNs to fluid systems remains unclear, particularly for atmospheric or oceanic data assimilation.

In this study, we propose a novel data-assimilation scheme, four-dimensional SRDA (4D-SRDA), incorporating SR-mixup---a novel domain-generalization technique. In 4D-SRDA, DA and spatio-temporal SR are performed simultaneously by a NN, while the system’s time evolution is obtained from physics-based LR simulations. This scheme does not require ensemble members and is computationally efficient. To mitigate the risk of domain shifts and the associated deterioration in inference accuracy, we employed SR-mixup---a data-augmentation technique based on the mixup method \cite{Zhang+2018ICLR, Yao+2022NIPS}. The validity of this scheme was demonstrated using an idealized barotropic ocean jet \cite{David+2017OM}. The 4D-SRDA and SR-mixup are applicable to physics-based models that use gridded input and output data. This study implies the potential of SR and domain generalization, in the field of DA, for integrating physics-based and data-driven models.

The remainder of the paper is organized as follows: 4D-SRDA is first proposed in Section \ref{sec:4d-srda}. SR-mixup is then independently presented in Section \ref{sec:sr-mixup}. Numerical experiments are described in Section \ref{sec:methods}, where the NN in the 4D-SRDA system is trained using SR-mixup. Section \ref{sec:results} evaluates the performance of the 4D-SRDA system and discusses the robustness of inference enhanced by SR-mixup. Section \ref{sec:conclusions} presents the conclusions.

\section{\label{sec:4d-srda}Four-Dimensional Super-Resolution Data Assimilation (4D-SRDA)}

\subsection{\label{subsec:general-concept}General Concept}

4D-SRDA comprises a NN and a physics-based model. The latter is typically an atmosphere or ocean model and is simply referred to as a ``fluid model.'' The system’s time evolution is calculated from LR simulations using the fluid model. The DA and spatio-temporal SR are performed simultaneously using the NN. Unlike EnKF and 4D-Var, the proposed method does not require ensemble members or iterative calculations and is computationally efficient.

\begin{figure}[t]
    \includegraphics[width=\textwidth]{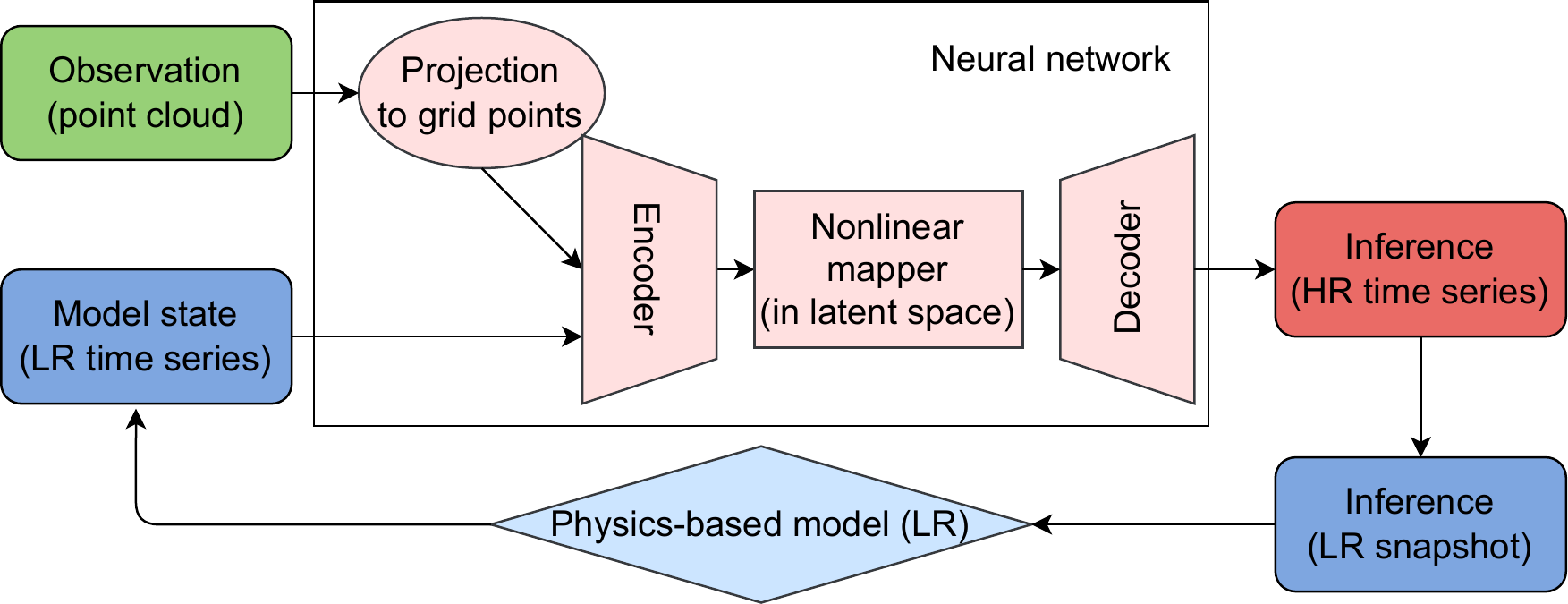}
    \caption{\label{fig:srda-cycle} Calculation process of four-dimensional super-resolution data-assimilation (4D-SRDA). The physics-based model, typically an atmosphere or ocean model, uses grid-point data as its input and output.}
\end{figure}

The calculation process in each assimilation cycle is shown in Figure \ref{fig:srda-cycle}. Consider the time series of observations and fluid-model states. The former are time-stamped and geolocated point-cloud data, while the latter are LR grid-point data. The LR fluid-model states generally include past, present, and future times, whereas observations are available only for the past and present. Using these two input types, the NN infers a flow-field time series at high spatio-temporal resolution. In the NN, the observations are converted from point-cloud to grid-point data. The converted observations and the LR model states are fused through a low-dimensional latent space by utilizing an encoder--decoder architecture. Because of the encoder, the observations do not necessarily have to be physical quantities---they can also be arbitrary quantities correlated with the true state of the system. In principle, the nonlinear mapper in Figure \ref{fig:srda-cycle} represents any mapping in the latent space, including temporal SR (Section \ref{subsec:neural-network}), while its specific form is determined by training. A snapshot of the inferred HR states is downsampled to the LR grid space using an algebraic operation, such as linear interpolation. The obtained LR snapshot is regarded as the analysis (the initial condition), from which an LR fluid simulation is conducted. The obtained output---a time series of LR model states---is used as input to the NN in the next assimilation cycle. This process (Figure \ref{fig:srda-cycle}) is repeated once new observations are obtained.

There are two key differences between 4D-SRDA and SRDA \cite{Barthelemy+2022OD}. First, in 4D-SRDA, the input to the NN contains a time series of fluid-model states, whereas in SRDA, the input is a snapshot of these states. Second, in 4D-SRDA, the NN simultaneously performs DA and spatio-temporal SR, whereas in SRDA, the NN conducts spatial SR followed by a statistical DA (e.g., EnKF). Both 4D-SRDA and SRDA obtain the time evolution of the system via LR fluid simulation. SRDA is easily applicable to existing DA frameworks because it performs SR and DA separately; however, this separation may not fully optimize SR inference for DA. In contrast, 4D-SRDA may achieve more accurate inference because it adopts end-to-end learning that simultaneously optimizes the SR and DA processes; however, the feasibility of such learning is unclear. This \change{paper}{study} presents an offline supervised learning method\add{ employing a new technique, SR-mixup,} \change{that}{which} mitigates the influence of domain shift\add{ (Section 3)}. Unsupervised or online learning is another important topic for future research.

\subsection{\label{subsec:inference-4d-srda}Algorithm of 4D-SRDA Inference}

\begin{figure}[t]
    \includegraphics[width=\textwidth]{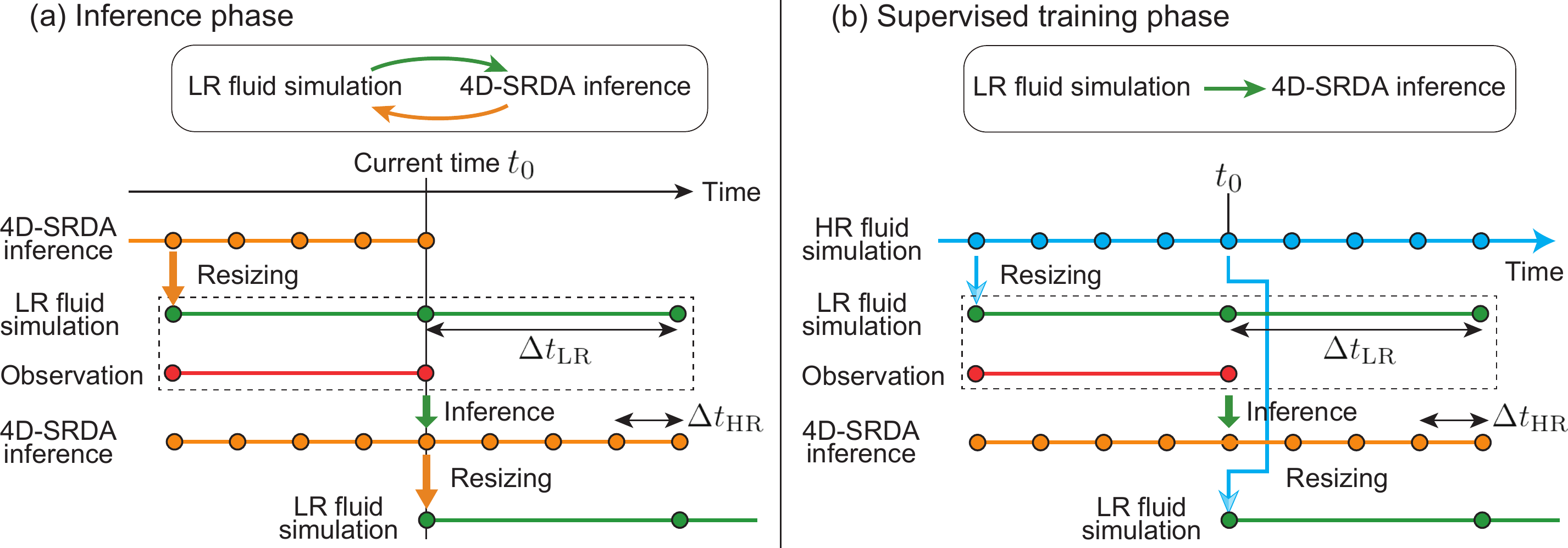}
    \caption{\label{fig:inference-train-method} Schematics for (a) inference and (b) supervised training. The low- and high-resolution (LR and HR) time steps are denoted by $\Delta t_{\rm LR}$ and $\Delta t_{\rm HR}$, respectively, where $\Delta t_{\rm LR} = 4 \Delta t_{\rm HR}$. The assimilation interval is assumed to be $\Delta t_{\rm LR}$. In (a), the 4D-SRDA inference and LR fluid simulation are alternated, whereas in (b), the 4D-SRDA inference is not used to conduct LR fluid simulations. Thus, (b) does not consider feedback from the neural network to the LR fluid model. The configuration illustrated here was employed in the numerical experiments in Sections \ref{sec:methods} and \ref{sec:results}.}
\end{figure}

The details of the 4D-SRDA inference are given for one assimilation cycle shown in Figure \ref{fig:inference-train-method}a, where the assimilation window has the same length as the forecast period. This configuration is used for the numerical experiments in Sections \ref{sec:methods} and \ref{sec:results}. More general cases are briefly discussed later in this section.

We first describe the input to the NN. A time series of LR states is obtained by numerically integrating the LR fluid model:
\begin{equation}
    \{\bm{X}^{LR}(t_0 + i \Delta t_{\rm LR})\}_{i=-1,0,1} = \{ \bm{X}^{\rm LR}(t_0 - \Delta t_{\rm LR}), \bm{X}^{\rm LR}(t_0), \bm{X}^{\rm LR}(t_0 + \Delta t_{\rm LR}) \}, \label{eq:time-sereis-LR-states}
\end{equation}
where the current time is denoted by $t_0$ and $\Delta t_{\rm LR}$ is the LR time step for the output. The initial condition $\bm{X}^{\rm LR}(t_0 - \Delta t_{\rm LR})$ is given by the previous analysis state; $\bm{X}^{\rm LR}(t_0)$ and $\bm{X}^{\rm LR}(t_0 + \Delta t_{\rm LR})$ are the current and future states estimated by the LR fluid model, respectively. Each snapshot $\bm{X}^{\rm LR}$ is defined on the LR spatial grid points and is generally composed of various three-dimensional field quantities, such as velocity and temperature. Observations are assumed to be given as:
\begin{equation}
    \{\bm{O}^{\rm HR}(t_0 + j \Delta t_{\rm LR})\}_{j=-1,0} = \{ \bm{O}^{\rm HR}(t_0 - \Delta t_{\rm LR}), \bm{O}^{\rm HR}(t_0) \}, \label{eq:time-series-observations}
\end{equation}
which does not include the future time $t_0 + \Delta t_{\rm LR}$. The observation $\bm{O}^{\rm HR}$ also consists of several three-dimensional quantities and is defined on the HR spatial grid points, including missing data. The HR spatial resolution is assumed to be four times finer than the LR resolution.

A trained NN, denoted by $f$, super-resolves the time series of the LR fluid-model states while incorporating the observations:
\begin{equation}
    \{\bm{\hat{Y}}^{HR}(t_0 + k \Delta t_{\rm HR})\}_{k=-4,\dots,4} = f\left(\{\bm{X}^{LR}(t_0 + i \Delta t_{\rm LR})\}_{i=-1,0,1},  \{\bm{O}^{\rm HR}(t_0 + j \Delta t_{\rm LR})\}_{j=-1,0}\right). \label{eq:inference-4dsrda}
\end{equation}
The time series of $\bm{\hat{Y}}^{\rm HR}$ is composed of the same three-dimensional quantities as $\bm{X}^{\rm LR}$, whereas $\bm{\hat{Y}}^{\rm HR}$ is obtained on the HR spatio-temporal grid points. For instance, in Figure \ref{fig:inference-train-method}a, the HR temporal resolution is four times finer, i.e., $\Delta t_{\rm HR} = \Delta t_{\rm LR}/4$. A hat indicates that $\bm{\hat{Y}}^{\rm HR}$ is a NN inference. \add{The input of the future LR state $\bm{X}^{LR}(t_0 + \Delta t_{\rm LR})$ is expected to serve as auxiliary information for time evolution, helping the NN infer the HR states at future times. }The output $\bm{\hat{Y}}^{\rm HR}(t_0)$ for the current time $t_0$ is mapped onto the LR spatial grid points using an algebraic method, such as linear interpolation. This mapped state is utilized as the initial condition for the LR fluid model in the next assimilation cycle, in which the current time $t_0$ is moved to $t_0 + \Delta t_{\rm LR}$ (the assimilation interval is $\Delta t_{\rm LR}$).\add{ The outputs at the other time steps, such as $\bm{\hat{Y}}^{\rm HR}(t_0-\Delta t_{\rm HR})$ and $\bm{\hat{Y}}^{\rm HR}(t_0+\Delta t_{\rm HR})$, are not used in the next assimilation cycle.}

Various modifications of the above algorithm are possible. For instance, more LR time steps can be included in the inference, whereas three time steps are used in Equation \ref{eq:inference-4dsrda}; the length of the assimilation window or forecast period can be varied, whereas both have the same length in Figure \ref{fig:inference-train-method}a; the scale factor of the spatial or temporal SR can be changed, whereas both factors are set to four above; and the past data can be eliminated \change{in the inference, whereas the LR state and observations in the past ($t_0 - \Delta t_{\rm LR}$) are included in}{from the input to the NN, leaving only the present and future input data, in contrast to} Equation \ref{eq:inference-4dsrda}.

In this research, two assumptions are made for observation data: (i) mappability to any space-time point and (ii) statistical stationarity. Equation \ref{eq:time-series-observations} is based on the first assumption; that is, \change{by using spatio-temporal interpolation, snapshots of the observations can be obtained at the LR time intervals and mapped to HR spatial grid points.}{the observations can be mapped to the HR spatial grid points and LR time points by spatio-temporal interpolation.} In real-world applications, \change{the interpolation or mapping of observations is not easily performed; for instance, it is difficult to map surface precipitation directly to four-dimensional space-time coordinates.}{such interpolation cannot be directly applied to integrated quantities. For instance, mapping cumulative precipitation to the time axis is not straightforward.}\add{ Investigating the treatment of integrated quantities is beyond the scope of this study. Thus, we make the first assumption.} NNs can be effective in solving such an inverse problem. In particular, the reconstruction techniques of three-dimensional point cloud data from two-dimensional images \cite<e.g.,>{Mandikal+2018BMVC, Li+2020IEEE} would be useful to assimilate integrated observations.

The second assumption covers stationarity in terms of both temporal and spatial statistics, which indicates that observation errors, including measurement and representativeness errors, are stationary. This assumption facilitates the representation learning of observations with missing data (Section \ref{subsec:obs-feature}), but is not necessarily satisfied in real-world applications. It is worth noting here that the statistical stationarity does not imply temporal steadiness. Indeed, in Sections \ref{sec:methods} and \ref{sec:results}, the spatial density of observations is constant, while the observed grid points vary in time. This change in observed locations is accounted for by the translation equivariance of the convolution operation \cite<e.g.,>{Cohen+2019NIPS}; i.e., features can be extracted by convolution independently of observed locations. This implies that the local equivariance improves the generalizability for changes in observation data. For instance, there may be cases where the stationarity is broken globally but preserved locally, such as in the expansion of observation networks. Various equivariant NNs have recently been studied \cite<e.g.,>{Cohen+2019NIPS, Sosnovik2021}, including graph NNs \cite{Fuchs2020}. Such an advanced architecture will need to be incorporated into 4D-SRDA for practical applications where observations are not stationary.

\subsection{\label{subsec:training-4d-srda}A Case of Supervised Learning}

\change{In 4D-SRDA, the NN is trained using a certain method. This study presents an offline supervised learning method.}{In this study, the NN involved in 4D-SRDA is trained using an offline supervised learning method.} Offline means that training data are generated in advance from LR fluid simulations in which the NN inference is not considered; in other words, the feedback from the NN to the LR fluid model is not considered in the offline training. Supervised learning requires HR target data that can be interpreted as the true state of the atmosphere or ocean. In practice, such data can be obtained from HR reanalysis datasets. Thus, in the case of supervised learning, 4D-SRDA is a computationally efficient alternative for generating the HR reanalysis.

Figure \ref{fig:inference-train-method}b describes the supervised learning that corresponds to the inference in Figure \ref{fig:inference-train-method}a. Consider the target variables, namely the HR time series $\bm{Y}^{\rm HR}(k \Delta t_{\rm HR})$ $(k = 0, \dots, N)$, where each snapshot $\bm{Y}^{\rm HR}$ is defined on the HR spatial grid points. Training datasets are created at a constant interval of $\Delta t_{\rm LR}$, which is equal to the assimilation interval in the inference phase (Figure \ref{fig:inference-train-method}a). Here, we explain the generation of a \add{sample of the }dataset for reference time $t_0$. The initial condition for the LR fluid model is given by resizing the HR snapshot $\bm{Y}^{\rm HR}(t_0 - \Delta t_{\rm LR})$ by an algebraic method, such as linear interpolation. This initial condition is interpreted as the analysis state from the previous assimilation cycle. The LR fluid model is then numerically integrated from $t_0 - \Delta t_{\rm LR}$ to $t_0 + \Delta t_{\rm LR}$ to obtain a time series of the LR states $\bm{X}^{\rm LR}$, as in Equation \ref{eq:time-sereis-LR-states}. For observations, since the target variables \change{can be}{are} obtained from \change{a}{the HR} reanalysis dataset, \change{the observations assimilated to this dataset}{the same observations used to produce this reanalysis} can be utilized in the training. By using spatio-temporal interpolation, the observations $\bm{O}^{\rm HR}$ are generated at $t_0 - \Delta t_{\rm LR}$ and $t_0$, as in Equation \ref{eq:time-series-observations}. Consequently, the pair of the input and output is given by
\begin{equation}
    \left[ \Bigl(\{\bm{X}^{LR}(t_0 + i \Delta t_{\rm LR})\}_{i=-1,0,1}, \{\bm{O}^{\rm HR}(t_0 + j \Delta t_{\rm LR})\}_{j=-1,0} \Bigr), \{\bm{Y}^{HR}(t_0 + k \Delta t_{\rm HR})\}_{k=-4,\dots,4} \right].
\end{equation}
Training data generation is repeated by varying $t_0$ at a constant interval of $\Delta t_{\rm LR}$. When optimizing the NN, the inference is computed as in Equation \ref{eq:inference-4dsrda}, and the NN weights are updated based on the gradients of a given loss function.

The generation of the initial condition for the LR fluid model is different between the inference and training phases (Figure \ref{fig:inference-train-method}). Specifically, in the inference phase, the initial condition is obtained from the previous NN inference, while in the training phase, it is obtained from the HR target data. This difference results in a domain shift  \cite<e.g.,>{MorenoTorres+2012PR}, namely a statistical difference in the input time series of LR states $\bm{X}^{\rm LR}$, which can be detrimental to 4D-SRDA performance. Indeed, a domain shift is inevitable in offline training because the training data are generated in advance independently of the NN inference. As a potential solution, we present a novel technique, SR-mixup.

\section{\label{sec:sr-mixup}Super-Resolution Mixup (SR-mixup)}

SR-mixup is a data augmentation method for domain generalization. The algorithm can be applied to ordinary SR problems; thus, the following discussion is independent of 4D-SRDA. Before explaining SR-mixup, we will review the mixup method \cite{Zhang+2018ICLR, Yao+2022NIPS}.

Mixup is a data augmentation method for supervised classification problems \cite{Zhang+2018ICLR}. Consider a training dataset $\{(X^{(k)}, Y^{(k)})\}_{k=1}^D$ ($D \in \mathbb{N}$), where $X^{(k)}$ is a numerical array, such as an image, and $Y^{(k)}$ is the corresponding class label based on one-hot encoding. A linear combination is considered for two randomly sampled pairs, $(X^{(i)}, Y^{(i)})$ and $(X^{(j)}, Y^{(j)})$:
\begin{eqnarray}
    \tilde{X} &=& \lambda X^{(i)} + (1 - \lambda) X^{(j)}, \label{eq:mixup-x} \\
    \tilde{Y} &=& \lambda Y^{(i)} + (1 - \lambda) Y^{(j)}, \label{eq:mixup-y}
\end{eqnarray}
where the interpolation ratio $\lambda \in [0, 1]$ is drawn from a beta distribution $B(a,b)$ ($a, b \in \mathbb{R}^+$). Supervised learning is performed for the mixed sample $(\tilde{X}, \tilde{Y})$. Mixup is applied only during training, and not during the test phase. The random linear combination generates synthetic samples with a different distribution from the original training data. This mixing is expected to enhance robustness to domain shifts.

Several variants of mixup have been proposed \cite<e.g.,>{Verma+2019PMLR, Mancini+2020ECCV, Yao+2022NIPS}, most of which are intended for classification. \citeA{Yao+2022NIPS} proposed a mixup for regression, C-Mixup, to address the issue that the linearity imposed via mixup reduces the accuracy of regression models when data are highly nonlinear. In C-Mixup, the linear combination in Equations \ref{eq:mixup-x} and \ref{eq:mixup-y} is created for similar randomly sampled pairs, where the similarity is measured using a distance metric between the labels $Y^{(i)}$ and $Y^{(j)}$.

SR-mixup is based on C-Mixup. Super-resolution problems are treated as regression when SR models are trained via supervised learning \cite<e.g.,>{Dong+2014ECCV}. In such cases, the input $X$ is an LR image and the output $Y$ is the corresponding HR image. The linear combination of similar HR images results in blurred images. Because the purpose of SR is to improve the resolution and clarity of LR images, C-Mixup may not be suitable for training SR models. We propose applying C-Mixup only to the input LR image. This method is called SR-mixup to emphasize the mixup for SR. Similarity is determined from the input $X$ instead of $Y$. SR-mixup is applicable to both supervised and unsupervised learning because the pairing between $X$ and $Y$ is not required.

The SR-mixup algorithm consists of four steps and is employed only during training. Here, we explain the algorithm in a supervised setting, which is similar for unsupervised applications. First, a pair $(X^{(i)}, Y^{(i)})$ is drawn from the training dataset. Second, $n$ other pairs are randomly sampled, and pair $(X^{(j)}, Y^{(j)})$ is further extracted such that the Euclidean distance between $X^{(i)}$ and $X^{(j)}$ is the minimum. We adopted Euclidean distance as the dissimilarity measure, although other metrics can be used. Third, an interpolation ratio $\lambda$ is drawn from the beta distribution $B(a,b)$. Fourth, a linear combination of inputs is created, $\tilde{X}^{(i)} = \lambda X^{(i)} + (1 - \lambda) X^{(j)}$, resulting in the augmented pair $(\tilde{X}^{(i)}, Y^{(i)})$. The above four steps are repeated to generate a batch of data for optimizing the NN. The pairing of $X$ and $Y$ is not essential, because the augmented input $\tilde{X}^{(i)}$ is produced independently from the labels $Y^{(i)}$ and $Y^{(j)}$. The parameters were set to $n = 20$ for the random sampling in the second step and $a = b= 2$ for the beta distribution $B(a,b)$ in the third step. Note that these parameters are context specific. In the present setting, $X^{(i)}$ and $X^{(j)}$ are regarded as statistically symmetric based on the symmetry of the beta distribution $B(a,b)|_{a=b=2}$. This symmetry is discussed in Section \ref{subsec:train-test-methods}.

\section{\label{sec:methods}Methods}

Numerical experiments were performed to confirm the effectiveness of 4D-SRDA and SR-mixup. Specifically, we simulated idealized barotropic ocean jets and then conducted supervised learning. Hereafter, 4D-SRDA is referred to as ST-SRDA (spatio-temporal SRDA) because the barotropic jet was modeled as a two-dimensional flow and the sum of the spatial and temporal dimensions is not four.

\subsection{\label{subsec:fluid-simulation} Fluid Simulation}

An idealized barotropic ocean jet was simulated in a two-dimensional periodic channel \cite{David+2017OM}. The equatorial beta plane $(x,y)$ was utilized, where $x \in [0, 2\pi]$ and $y \in [0, \pi]$ are the zonal and meridional coordinates, respectively. The governing equations are as follows: 
\begin{eqnarray}
  \frac{\partial \omega}{\partial t} + u \frac{\partial \omega}{\partial x} + v \frac{\partial \omega}{\partial y} + \beta v &=& - r \omega - \nu \Delta^2 \omega - \frac{{\rm d}\tau(y)}{{\rm d}y}, \label{eq:vorticity-evolution} \\
  \Delta \psi &=& \omega, \label{eq:poisson-eq} \\
  \left(u, v\right) &=& \left(-\frac{\partial \psi}{\partial y}, \frac{\partial \psi}{\partial x}\right). \label{eq:velocity-def}
\end{eqnarray}
Equation \ref{eq:vorticity-evolution} describes the evolution of vorticity $\omega$ with respect to time $t$, where the left-hand side represents the advection of $\omega$ and the beta effect, and the right-hand side consists of linear drag, hyperviscosity, and forcing due to zonal wind stress $\tau(y)$. Equations \ref{eq:poisson-eq} and \ref{eq:velocity-def} give the stream function $\psi$ and velocity $(u,v)$, respectively. Wind stress $\tau(y)$ is given by
\begin{equation}
    \tau(y) = \tau_0 \; \left[ {\rm sech}^2\left(\frac{y - y_0}{\delta}\right) - c \right],
\end{equation}
where $c$ is determined such that the integral of $\tau(y)$ is zero. The parameters in the governing equations were set as follows: $\beta = 0.1$, $r = 1 \times 10^{-2}$, $\nu = 1 \times 10^{-5}$, $\tau_0 = 0.3$, $y_0 = \pi/2$, and $\delta = 0.4$. According to \citeA{David+2017OM}, this configuration is within the parameter regime of mixing barriers with strong eddies, where coherent vortices remain in a statistically steady state (Section \ref{subsec:accuracy-srda}). The presence of these vortices simplifies the evaluation because the phase of vorticity, namely the large-scale flow structure, is related to errors in inference.

The initial condition was a superposition of a zonal jet and random perturbations. The zonal jet $U(y)$ is given by 
\begin{equation}
    U(y) = U_0 \; \left[ {\rm sech}^2\left(\frac{y - y_0}{\delta}\right) - c \right], \label{eq:zonal-jet}
\end{equation}
where $U_0 = 3.0$. This initial jet satisfies the Rayleigh-Kuo inflection-point criterion, i.e., $\beta - \frac{d^2 U(y)}{dy^2}$ being $0$ somewhere in the domain, which is a necessary condition for barotropic instability. The perturbations were added to the vorticity field for each wavenumber, where the phase and amplitude were sampled from Gaussian distributions with means of $0$ and standard deviations of $\pi$ for the phase and $2.5 \times 10^{-3}$ for the amplitude. The simulation data for deep learning were generated by varying the random perturbations.

Numerical integration was performed using the modified Euler and pseudo-spectral methods. Three different resolutions were adopted: LR, HR, and ultra-HR (UHR). All configurations are summarized in Table \ref{table:cfd-simulation-config}. The integration time was from $t = 0$ to $24$ for all experiments.

\begin{table}[t]
    \caption{\label{table:cfd-simulation-config} Fluid simulation configurations for three different resolutions.}
    \centering
    \begin{tabular}{l r r r r}
        \hline
        Name  & \multirow{2}{4em}{Grid size ($x \times y$)} & \multirow{2}{*}{\change{Truncation}{Cutoff} wavenumber} & \multirow{2}{5em}{Integration time step} & \multirow{2}{5em}{Output time step}  \\ \\
        \hline
        LR (low resolution) & 32 $\times$ 16  & \remove{T}10 & $5 \times 10^{-4}$ & 1.00 ($= \Delta t_{\rm LR}$) \\
        HR (high resolution) & 128 $\times$ 64 & \remove{T}42 & $\frac{1}{4} \times 5 \times 10^{-4}$ & 0.25 ($= \Delta t_{\rm HR}$) \\
        UHR (ultra-high resolution) & 1024 $\times$ 512 & \remove{T}341 & $\frac{1}{32} \times 5 \times 10^{-4}$ & 0.25 ($= \Delta t_{\rm HR}$) \\
        \hline
    \end{tabular}
\end{table}

\subsection{\label{subsec:neural-network} Neural Network in the ST-SRDA System}

The calculation process in the NN is simplified for potential application to three-dimensional flows. In particular, the temporal SR is separated from the spatial SR; that is, the temporal SR is performed in the nonlinear mapper, while the spatial SR is conducted in the decoder. In addition, the encoder and decoder act independently on the state at each time.

\begin{figure}[t]
    \includegraphics[width=\textwidth]{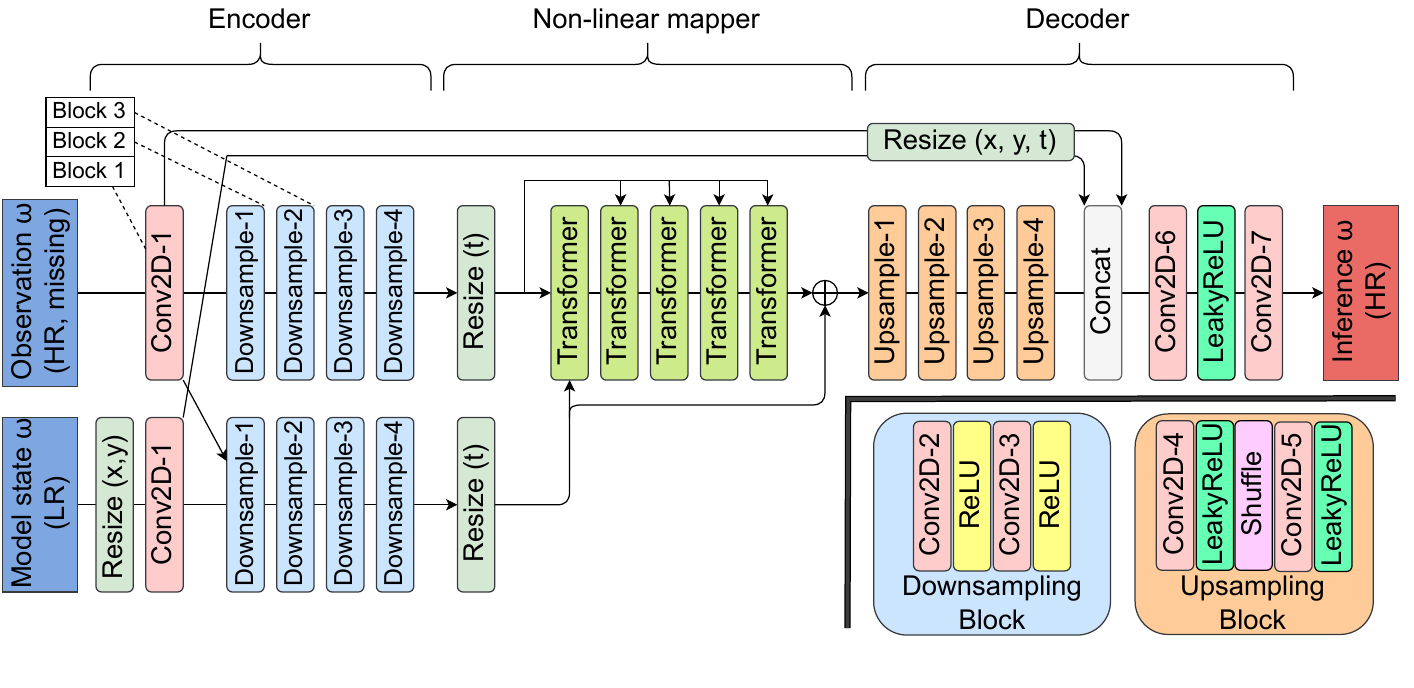}
    \caption{\label{fig:model-architecture} Neural network architecture of the spatio-temporal SRDA (ST-SRDA) system. The symbol $\omega$ denotes vorticity fields. The blocks ``Downsample'' and ``Upsample'' are described in the bottom right corner. In the upsampling block, the label ``Shuffle'' refers to the pixel shuffle layer \cite{Shi+2016CVPR}. If two or three arrows point to a block, all inputs are concatenated and then fed to the first layer in the block. Details about the implementation are available at the Zenodo repository (see Open Research). Outputs from Block 1 through 3 are shown in Figure \ref{fig:obs-feature}.}
\end{figure}

Figure \ref{fig:model-architecture} outlines the network architecture, which is based on U-Net \cite{Ronneberger+2015MICCAI}. U-Net architectures have been employed in the SR for fluid systems \cite<e.g.,>{Jiang+2020ICHPC, Wang+2021PF, Hammoud+2022JAMES}. The input and output of the NN are vorticity fields. Compared with Figure \ref{fig:srda-cycle}, this NN does not include a transformation from point-cloud to grid-point data because the observations are defined on the HR spatial grid points (Equations \ref{eq:time-series-observations} and \ref{eq:inference-4dsrda}). Hyperparameters, such as the number of hidden layers, were tuned via a grid search. These parameters are presented in \ref{sec:hyper-param-nn}. Details on the implementation of the NN are available at the Zenodo repository (see Open Research).

The encoder consists of four downsampling blocks. Each block begins with a convolution operation with a $2 \times 2$ stride and halves the spatial size of the input array in the $x$ and $y$ dimensions. The LR model states are spatially resized\add{ using nearest neighbor interpolation} to match the HR \add{(``Resize (x,y)'' in Figure 3) }and then encoded while being combined with the observations. Additionally, the observations are independently encoded.

The nonlinear mapper contains five transformer-encoding blocks \cite{Vaswani+2017NIPS}. The time interval of the input features is refined via linear interpolation\add{ to match the HR time steps (``Resize (t)'' in Figure 3)}. This interpolation, followed by the nonlinear mapping, performs the temporal SR. The output from each transformer block is passed to the next, and the encoded observations are input to all blocks. \add{Each block first computes weights for the input features across every time step pair and then outputs the weighted sums of the features. This processing, known as the scaled dot-product attention, incorporates temporal correlations into the nonlinear mapping} \cite{Vaswani+2017NIPS, Niu+2021NC}. Finally, the encoded LR model states are added to the output via a skip connection. 

The decoder comprises four upsampling blocks. Each block utilizes pixel shuffle \cite{Shi+2016CVPR} to double the spatial size of the input array in the $x$ and $y$ dimensions. Using these four blocks, the spatial size of the output is restored to the HR grid size. The obtained output is concatenated with the features extracted by the first convolution in the NN\add{, which are resized using linear interpolation along the space and time axes (``Resize (x,y,t)'' in Figure 3)}. The final inference is given through the last three convolution and nonlinear layers.

In preprocessing, the vorticity at each grid point is transformed by the following formula:
\begin{equation}
    {\rm clip}_{[0, 1]}\left(\frac{\omega - m}{s}\right). \label{eq:scaling}
\end{equation}
The clipping function, defined as ${\rm clip}_{[0,1]}(z) = \min\{1, \max\{0, z\}\}$ ($z \in \mathbb{R}$), limits the value range to $[0, 1]$. This function acts independently on each element of the input array. Parameters $m$ and $s$ were determined such that 99.999\% of the vorticity values were within the range of $[0,1]$. As a result of this preprocessing, $0$ was considered a missing value and was substituted onto the HR grid points without observations.

\subsection{\label{subsec:train-test-methods} Training and Testing Methods for the Neural Network}

The training and testing methods for the NN are based on Sections \ref{subsec:training-4d-srda} and  \ref{subsec:inference-4d-srda}, respectively. The relationship for the data along the time axis is the same as in Figure \ref{fig:inference-train-method}, where $\Delta t_{\rm LR} = 1.00$ and $\Delta t_{\rm HR} = 0.25$. The assimilation interval is $\Delta t_{\rm LR}$, which is approximately half of the advection time scale $2\pi / U_0$.

The training was performed using the HR simulation data, whereas in the test phase, ST-SRDA performance was evaluated with the UHR simulation results. The ST-SRDA is a surrogate scheme for generating HR reanalysis in the case of supervised learning (Section \ref{subsec:training-4d-srda}). In the training phase, the HR simulation data are considered to be derived from a reanalysis, while in the test phase, the UHR data are interpreted as the true state of the barotropic ocean jet system. The spatio-temporal resolution of the ST-SRDA output is always HR (Table \ref{table:cfd-simulation-config}); in the test, performance was measured by mapping the HR inference to UHR via bicubic interpolation. For comparison, EnKFs were applied to the LR and HR fluid models. The latter, EnKF at the HR, is interpreted as a model that generates the HR training data and is expected to provide a lower error bound for the ST-SRDA (The EnKFs are described in Section \ref{subsec:enkf}).

The scale factor for the SR is $4$ in both the spatial and temporal dimensions (i.e., $x$, $y$, and $t$). The target variable is a part of the HR vorticity time series, comprising nine data points with a time step of $\Delta t_{\rm HR}$ ($= 0.25$) (Figure \ref{fig:inference-train-method}b). The first four points are interpreted as the past, the middle point the present, and the last four points the future. The corresponding inputs to the NN are the time series of the observations and LR fluid-model states. The observations consist of two points in time, whereas the LR states consist of three points (Figure \ref{fig:inference-train-method}b). The time interval of both input time series is $\Delta t_{\rm LR}$ ($= 1.0$). This configuration means that the NN performs the temporal SR with a factor of $4$. Since the LR spatial grids ($32 \times 16$) are four times coarser than the HR grids ($128 \times 64$), the spatial SR also has a factor of $4$. The scale factor of $4$ is sufficiently high for real applications \cite{Onishi+2019SOLA, Wang+2021GMD}.

Synthetic observations were generated from the HR and UHR simulations in the training and test phases, respectively. For HR, observed grid points were randomly subsampled with a constant spatial spacing. Specifically, the spatial density of observations was \remove{set to a }constant\remove{ (0.5--6.5\%)}, while observed grid points varied in time. For instance, when the vorticity is observed at every $8 \times 8$ grids, the spatial density is approximately 1.56\%. This ratio is referred to as the observation point ratio, which is defined as
\begin{equation}
    \frac{\text{the total number of observed grid points}}{\text{ the total number of HR grid points $(= 128 \times 64)$ }}. \label{eq:def-obs-ratio}
\end{equation}
\add{The sensitivity to this ratio was assessed across experiments by varying it from 0.5\% to 6.5\%. }At the unobserved grid points, Not a Numbers (NaNs) were substituted and replaced in subsequent preprocessing with missing values, i.e., zeros (Section \ref{subsec:neural-network}). Spatially independent Gaussian noise was added to the subsampled vorticity, with a mean of $0$ and a standard deviation of $0.1$, which is approximately 5\% of the spatio-temporal average of the absolute values of vorticity. For UHR, the average pooling of an $8 \times 8$ kernel was applied to the vorticity field to match the spatial size to that of the HR, and the above processing was applied.

The training and validation data were generated from the HR fluid simulations (Figure \ref{fig:inference-train-method}b). All simulations were terminated at $t = 24$, \change{where}{when} the barotropic jet becomes a statistically steady state consisting of westward-propagating large-scale eddies \cite{David+2017OM}. To increase dataset size, the initial random perturbations were varied randomly (Section \ref{subsec:fluid-simulation}), and 3,500 and 1,000 simulations were run for the training and validation dataset, respectively. Each HR simulation contained \change{22}{23} LR-simulation cycles because of $\Delta t_{\rm LR} = 1$ (i.e., \change{$t_0 = 2, \dots, 23$}{$t_0 = 1, \dots, 23$}). In each cycle, the initial condition for the LR fluid model was given by applying a low-pass filter to the HR vorticity in the wavenumber domain. The total size of the training and validation dataset was \change{77,000 ($= 3,500 \times 22$)}{80,500 ($= 3,500 \times 23$)} and \change{22,000 ($= 1,000 \times 22$)}{23,000 ($= 1,000 \times 23$)}, respectively.

The NN was trained by employing SR-mixup and Adam optimization \cite{Kingma+2015ICLR}. For the loss function, we used the mean absolute error (MAE) of vorticity. The weights of the NN that resulted in the smallest validation MAE were stored after 1,000 epochs with a mini-batch size of 32. SR-mixup was applied only to the LR model states. SR-mixup is not necessary for the observation data, because they are independent of the NN inference and not subject to domain shifts. If SR-mixup were applied to data with missing values, such as observations, the linear interpolation would need to be replaced with a nearest neighbor interpolation using a threshold against $\lambda$.

In each iteration, SR-mixup generates a synthetic time series as in Equation \ref{eq:mixup-x}:
\begin{eqnarray}
    \tilde{\omega}^{\rm LR, (i)}(t) &=& \lambda {\omega}^{{\rm LR}, (i)}(t) + (1 -\lambda) {\omega}^{{\rm LR}, (j)}(t), \notag \\ 
    &=& {\omega}^{{\rm LR}, (i)}(t) + (1 -\lambda) ({\omega}^{{\rm LR}, (j)}(t) - {\omega}^{{\rm LR}, (i)}(t)), \label{eq:mixup-vorticity} \\
    &\quad& \quad\quad\quad\quad\quad\quad\quad\quad\; \left(t \in \left\{t_0 - \Delta t_{\rm LR}, t_0, t_0 + \Delta t_{\rm LR}\right\}\right), \notag
\end{eqnarray}
where $i$ and $j$ denote the data indices, and $\lambda$ is independent of $t$ and sampled from the beta distribution $B(a,b)|_{a=b=2}$. Equation \ref{eq:mixup-vorticity} is interpreted as the original time series ${\omega}^{{\rm LR}, (i)}(t)$ being perturbed by a time series ${\omega}^{{\rm LR}, (j)}(t)$ that is similar to ${\omega}^{{\rm LR}, (i)}(t)$ in terms of Euclidean distance (Section \ref{sec:sr-mixup}). In training, $\tilde{\omega}^{\rm LR, (i)}(t)$ was used as input instead of ${\omega}^{{\rm LR}, (i)}(t)$. The beta distribution $B(a,b)$ was set to be symmetric (i.e., $a=b=2$) to increase the perturbation amplitude. The grid-wise relative error between two similar LR time series was found to be quite small ($\lesssim$ 10\%) due to the sufficiently large dataset size. As the dataset size decreases, the difference between ${\omega}^{{\rm LR}, (i)}$ and ${\omega}^{{\rm LR}, (j)}$ tends to increase, suggesting that $\lambda$ needs to be biased toward $1$, i.e., an asymmetric beta distribution is required to suppress the perturbation amplitude. Note that SR-mixup partially breaks the dynamical consistency in the LR time series; for instance, nonlinear conserved quantities, such as energy, are not constant in time. On the other hand, the linearity of SR-mixup indicates that linear constraints, such as Equations \ref{eq:poisson-eq} and \ref{eq:velocity-def}, are preserved because all physical quantities are linearly superposed by using the same weight of $\lambda$. Mixup, including SR-mixup, is a simple method for domain generalization \cite{Zhang+2018ICLR, Yao+2022NIPS}. Advanced methods, such as generative adversarial networks (GANs) \cite<e.g.,>{Wang+2022IEEE, Zhou+2022IEEE} may improve dynamical consistency.

The test method is the same as the inference method in Figure \ref{fig:inference-train-method}a, which repeats the feedback cycle between the NN and fluid model. In the first cycle, the initial condition for the LR fluid model was given by a superposition of random perturbations and the zonal jet (Equation \ref{eq:zonal-jet}) because of no NN inference, while in later cycles, the initial condition was obtained by applying the low-pass filter to the NN inference. Since the initial perturbations in the first cycle were random, the vorticity evolution would be largely different even using the HR fluid model. Here, 50 UHR simulations were performed to generate the ground-truth data. The following results were not sensitive to the number of UHR simulations (similar results were found even for 5 UHR simulations).

ST-SRDA performance was evaluated using two metrics, namely pixel-wise accuracy and pattern consistency, for the time series that consists of the analysis and forecast values. The ST-SRDA output includes the past, present, and future (Figure \ref{fig:inference-train-method}a). All vorticity time series evaluated in Section \ref{sec:results} were created as follows: in each assimilation cycle, the inferences between $t_0$ and $t_0 + 3 \Delta t_{\rm HR}$ (i.e., four time points) were extracted; after all cycles, all extracted values were combined into one time series.

The MAE ratio measures pixel-wise errors as follows:
\begin{equation}
    \text{MAE ratio at } t = \frac{1}{N}\sum_{i=1}^{N} \frac{\sum_{\text{UHR space}} |\omega^{(i)}(t) - \hat{\omega}^{(i)}(t)|}{\sum_{\text{UHR space}} |\omega^{(i)}(t)|}, \label{eq:mae-ratio}
\end{equation}
where $N$ is the number of test simulations ($N = 50$), $\omega^{(i)}(t)$ is the ground-truth UHR vorticity in the $i$-th simulation, and $\hat{\omega}^{(i)}(t)$ is the corresponding HR inference of ST-SRDA. The spatial size was matched by resizing the HR inference to the UHR grid space via bicubic interpolation.

The consistency of the spatial patterns was estimated using the mean structural similarity index measure (MSSIM) \cite{Wang+2004IEEE}:
\begin{eqnarray}
    \text{MSSIM loss at }t &=& 1 - {\rm MSSIM}(t) \nonumber \\ &=& 1 - \frac{1}{N} \sum_i \sum_{\text{UHR space}} \frac{\left(2\mu^{(i)}(t) \hat{\mu}^{(i)}(t) + {\rm C_1}^2 \right) \left(2\gamma^{(i)}(t) + {\rm C_2}^2 \right)}{\left(\mu^{(i)}(t)^2 + \hat{\mu}^{(i)}(t)^2 + {\rm C_1}^2\right) \left( \sigma^{(i)}(t)^2 + \hat{\sigma}^{(i)}(t)^2 + {\rm C_2}^2 \right)}, \label{eq:mssim}
\end{eqnarray}
where ${\rm C_1} = 0.01$, ${\rm C_2} = 0.03$, $\mu^{(i)}(t)$ and $\sigma^{(i)}(t)^2$ are the mean and variance of the ground truth, respectively; $\hat{\mu}^{(i)}(t)$ and $\hat{\sigma}^{(i)}(t)^2$ are the corresponding quantities of the ST-SRDA inference; and $\gamma^{(i)}(t)$ is the covariance between the ground truth and inference. The variables in the summation are calculated locally by applying a Gaussian filter; for instance, when computing the second moment, the Gaussian filter is applied to each vorticity field after squaring the vorticity value at each grid point. We confirmed that MSSIM losses are not sensitive to the window function. Indeed, similar results were found when a rectangular window was employed instead of the Gaussian. The MSSIM loss takes a value greater than or equal to $0$, and smaller values indicate that the spatial patterns of inference are more similar to those of the ground truth. A detailed discussion of MSSIM can be found in \citeA{Wang+2004IEEE}.

The MAE ratio and MSSIM loss were calculated for each time $t$, as shown in Equations \ref{eq:mae-ratio} and \ref{eq:mssim}. The time-averaged values are referenced using the same names.

The evaluation in the test phase was conducted using an NVIDIA RTX A6000 GPU on a local workstation. The NN training was performed using four NVIDIA Tesla P100 GPUs on the TSUBAME3.0 supercomputer at the Tokyo Institute of Technology.

\subsection{\label{subsec:enkf} Ensemble Kalman Filter (EnKF)}

An EnKF \cite{Evensen1994JGRO} with the perturbed observation method \cite{Burgers+1998MWR} was applied to the HR or LR fluid model. The former, referred to as EnKF-HR, is considered to provide a lower bound on the errors for the ST-SRDA because the underlying HR fluid model is the same as that used to generate the training data. The EnKF applied to the LR fluid model is referred to as EnKF-SR, which is a baseline model for the ST-SRDA. In EnKF-SR, the assimilation is performed in the HR space after the LR fluid-model state is super-resolved to the HR by bicubic interpolation. \citeA{Barthelemy+2022OD} reported that performing the EnKF in the HR space may provide more accurate inferences than those in the LR space when observations are defined on the HR grid points. When the EnKF is applied in the LR space, several observations may need to be combined into a single point because of the coarser spatial resolution. The EnKF in the HR space does not require such aggregation, suggesting that the observation data are fully utilized. For both EnKF-HR and EnKF-SR, the initial condition for the fluid model was given by a superposition of random perturbations and the zonal jet (Equation \ref{eq:zonal-jet}) as in testing with the ST-SRDA, and the assimilation was performed at the constant interval $\Delta t_{\rm LR}$, i.e., the same interval of the ST-SRDA inference (Figure \ref{fig:inference-train-method}a). For simplicity, the observation errors, including representativeness errors, are assumed to obey the same statistics as described in Section \ref{subsec:train-test-methods} for all models (i.e., ST-SRDA, EnKF-SR, and EnKF-HR) because the assimilation is performed through the HR space in all models.

The background error covariance matrices were spatially localized using a function of \citeA{Gaspari+1999QJRMS}. These error covariances were inflated by adding Gaussian noise to the analysis state prior to the numerical integration of the fluid model \cite{Whitaker+2008MWR}. This Gaussian noise had a mean of $0$ and incorporated spatial correlation by estimating its covariance based on the training data at each time step. The initial perturbations to create ensemble members were also sampled from this Gaussian distribution at $t = 0$. The results were not sensitive to the Gaussian-noise covariance; similar results were obtained when the Gaussian noise had a constant correlation length. We also confirmed that this additive inflation provided more accurate inferences than the multiplicative covariance inflation \cite{Anderson+1999}

Hyperparameters for the EnKF-HR and EnKF-SR were determined for each observation point ratio such that the MAE was minimized using the training data. Specifically, the number of ensemble members was fixed at 100, and the following hyperparameters were tuned: the amplitude for the initial perturbation to create the ensemble; additive inflation amplitude; localization radius; and amplitude for perturbing observations.

\section{\label{sec:results}Results and Discussion}

\subsection{\label{subsec:accuracy-srda} Accuracy of ST-SRDA Inferences}

By comparing the EnKF-SR, EnKF-HR, and ST-SRDA, we will demonstrate that the ST-SRDA successfully achieves simultaneous DA and spatio-temporal SR. The test method repeats the feedback cycle between the fluid simulation and NN inference (Figure \ref{fig:inference-train-method}a) and is equivalent to the 4D-SRDA calculation process (Figure \ref{fig:srda-cycle}).

\begin{figure}[t]
    \includegraphics[width=\textwidth]{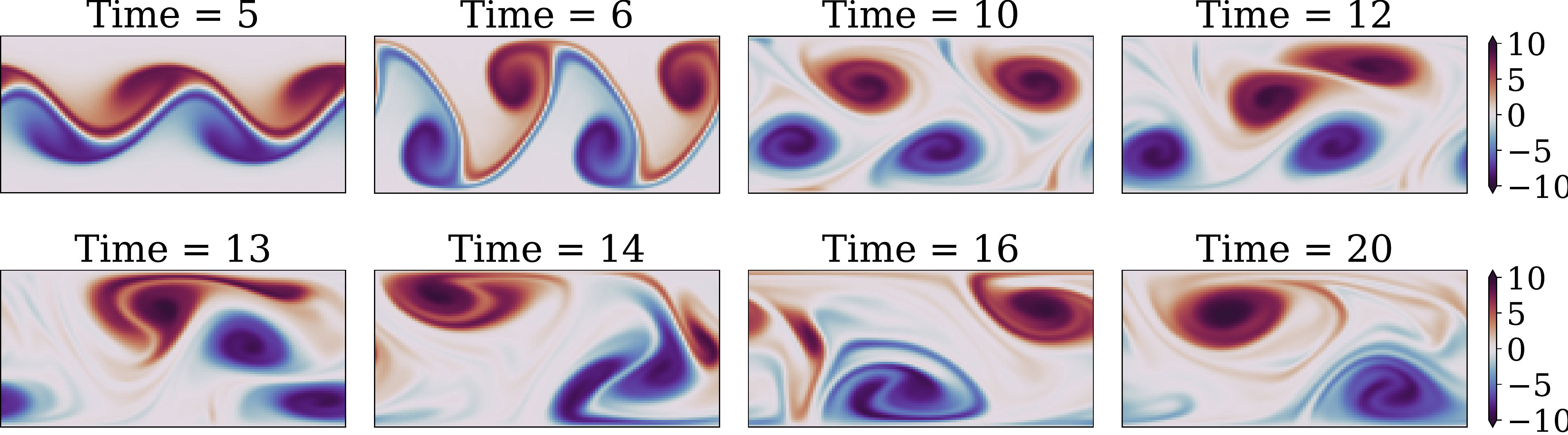}
    \caption{\label{fig:vorticity-evolution} An example of vorticity evolution from the ground-truth data at ultra-high resolution (UHR).}
\end{figure}

Before making this comparison, we discuss a typical vorticity evolution from the ground-truth UHR data (Figure \ref{fig:vorticity-evolution}). In the initial stage ($0 \le t \lesssim 10$), the unstable jet meanders and collapses into multiple vortices, while developing fine filaments. The vortices then merge into a flow with wavenumber 1 ($10 \lesssim t \lesssim 16$), where the small-scale structure is again enhanced ($t \sim 13$). Finally, the wavenumber-1 vorticity continues to propagate westward, i.e., in the negative $x$ direction ($16 \lesssim t$). This final state of the coherent vortices is characteristic of the regime of mixing barriers with strong eddies and can be understood through Rossby-wave dynamics \cite{David+2017OM}. All fluid simulations were terminated at $t=24$ (Section \ref{subsec:train-test-methods}) because no fine-scale pattern developed in this statistically steady state.

\begin{figure}[t]
    \includegraphics[width=\textwidth]{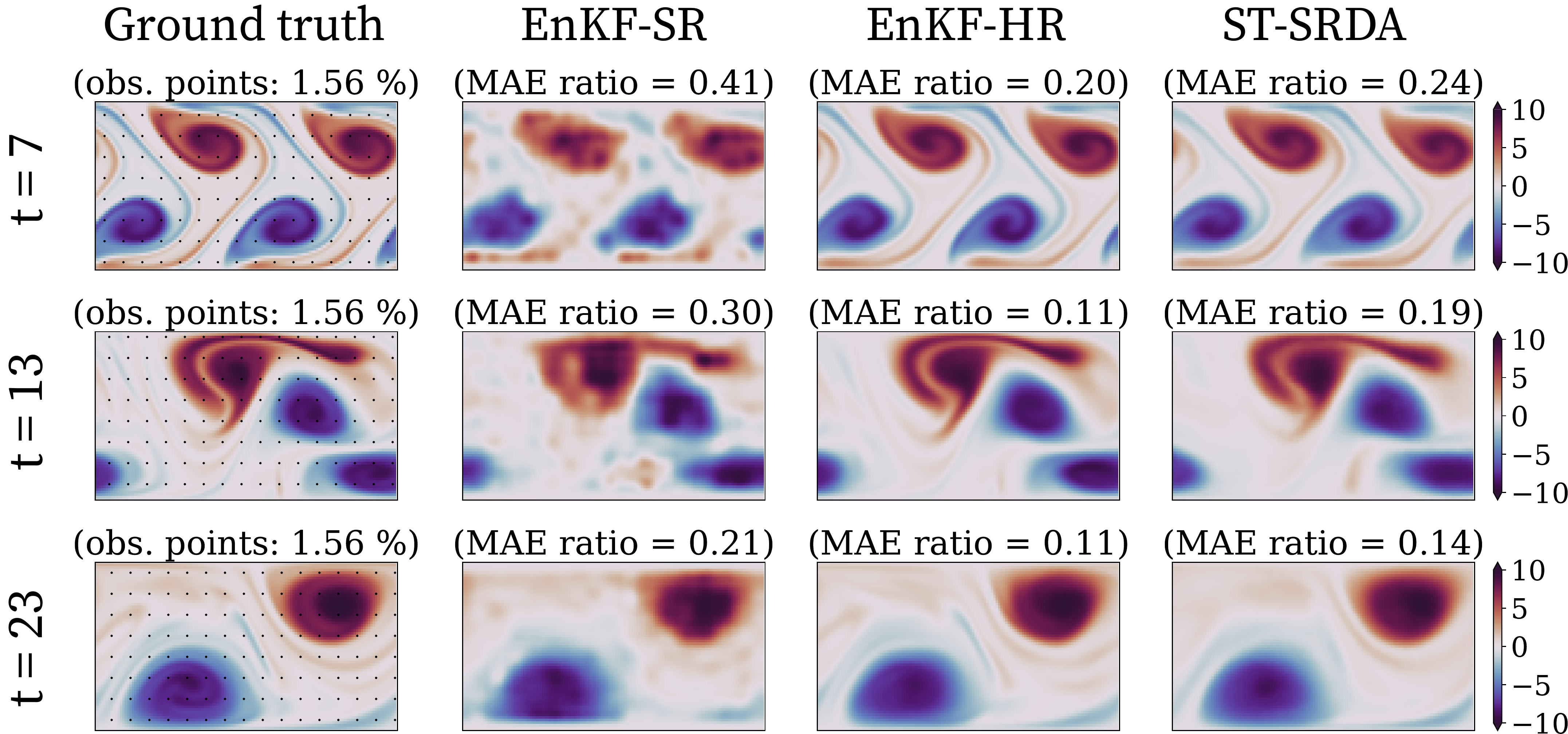}
    \caption{\label{fig:snapshots-enkf-srda} Vorticity snapshots from the ground-truth data and EnKF-SR, EnKF-HR, and ST-SRDA inferences. The black dots in the ground truth represent the observed grid points. Their ratio ($1.56$\%) was calculated according to Equation \ref{eq:def-obs-ratio}. The mean absolute error (MAE) ratio is defined by Equation \ref{eq:mae-ratio}.}
\end{figure}

The ST-SRDA successfully inferred both the small- and large-scale structures in the vorticity fields. Figure \ref{fig:snapshots-enkf-srda} compares the ground truth with the EnKF-SR, EnKF-HR, and ST-SRDA inferences. At $t=7$, the wavenumber-2 structure developed and was associated with the collapse of the jet stream (Figure \ref{fig:vorticity-evolution}). The ST-SRDA and EnKF-HR emulated this collapse with high accuracy; in particular, both reproduced the small-scale structures within the positive and negative vortices. Although the EnKF-SR inferred the vortex locations, it failed to reproduce the fine-scale patterns inside the vortices. At $t=13$, the results were similar to those at $t=7$: the vortex merging was successfully emulated by the ST-SRDA and EnKF-HR, but not by the EnKF-SR. At $t=23$, the large-scale coherent vortices propagated westward, and all models successfully inferred the phase of vorticity, i.e., the large-scale pattern. Overall, the ST-SRDA inferred vorticity fields that were similar to those of the EnKF-HR.

\begin{figure}[t]
    \includegraphics[width=\textwidth]{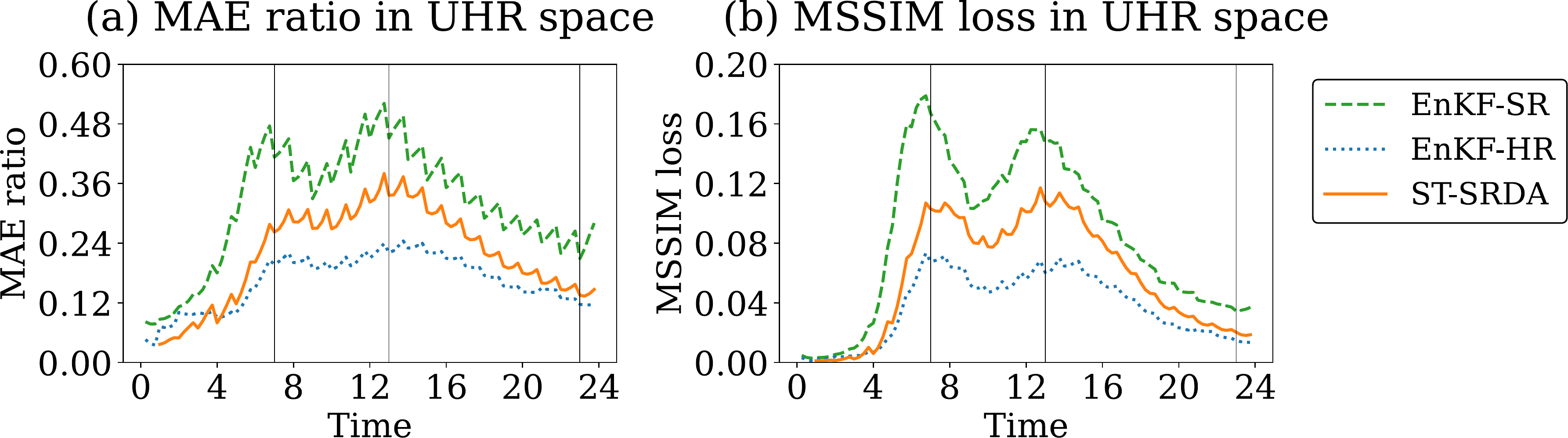}
    \caption{\label{fig:error-timeseries-enkf-srda} Time series of the test scores: (a) mean absolute error (MAE) ratio and (b) mean structural similarity index measure (MSSIM) loss. The MAE ratio and MSSIM loss are defined by Equation \ref{eq:mae-ratio} and \ref{eq:mssim}, respectively. The observation point ratio was set to $1.56$\%, as in Figure \ref{fig:snapshots-enkf-srda}. The vertical lines represent $t = 7$, $13$, and $23$. The inferences at these times are shown in Figure \ref{fig:snapshots-enkf-srda}. The assimilation interval is $\Delta t_{\rm LR} = 1.00$ for all the models.}
\end{figure}

The quantitative analysis indicated that the ST-SRDA inference tended to be more accurate than the EnKF-SR inference at all time points. Figure \ref{fig:error-timeseries-enkf-srda} shows the time series of the MAE ratio and MSSIM loss averaged over all of the test simulations. All scores were calculated in the UHR space, with LR and HR vorticity fields mapped to the UHR space via bicubic interpolation. The results were not sensitive to this mapping because a similar time series set was obtained when all of the scores were calculated in the LR or HR space. All time series in Figure \ref{fig:error-timeseries-enkf-srda} are based on the analysis and forecast values of the vorticity. Since the assimilation intervals ($\Delta t_{\rm LR}=1.00$) were the same for all models, the MAE ratio and MSSIM loss were reduced at the same time points. The time series of the ST-SRDA (orange solid lines) are missing for $0 \le t < 1$ because no forecast was inferred during this initial period (Section \ref{subsec:train-test-methods}). On average, the MAE ratio and MSSIM loss of the ST-SRDA were smaller than those of the EnKF-SR (green dashed lines), but larger than those of the EnKF-HR (blue dotted lines). This verifies that the EnKF-HR provides a lower bound on the errors for the ST-SRDA. The time series of the EnKF-SR have two peaks around $t=7$ and $13$ (vertical lines). These peaks reflect the development of fine-scale patterns due to jet stream collapse ($t \sim 7$) and vortex merging ($t \sim 13$) (Figures \ref{fig:vorticity-evolution} and \ref{fig:snapshots-enkf-srda}). These two peaks are observed for the MSSIM loss of the ST-SRDA and EnKF-HR (Figure \ref{fig:error-timeseries-enkf-srda}b), but are not clear for their MAE ratios (Figure \ref{fig:error-timeseries-enkf-srda}a). This result is likely due to the fact that the MAE ratio is a pixel-wise error and not sensitive to differences in flow patterns \cite<e.g.,>{Wang+2004IEEE}. After the vortex merging ($13 \lesssim t$), the MAE ratio and MSSIM loss decreased, while the flow field approached the statistically steady large-scale state. This result is plausible as it is relatively easy to infer large-scale flow patterns.

\begin{figure}[t]
    \includegraphics[width=\textwidth]{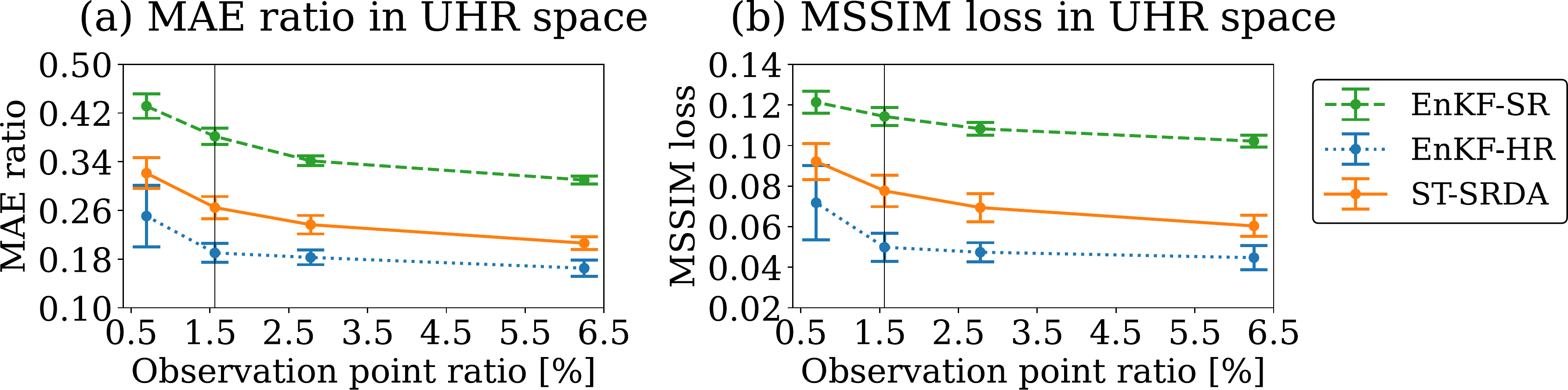}
    \caption{\label{fig:error-obs-ratio-enkf-srda} Dependence of the test scores on the observation point ratio: (a) mean absolute error (MAE) ratio and (b) mean structural similarity index measure (MSSIM) loss. The MAE ratio and MSSIM loss are defined by Equation \ref{eq:mae-ratio} and \ref{eq:mssim}, respectively. The test scores were averaged over $4 \le t \le 20$. The observation point ratio was calculated according to Equation \ref{eq:def-obs-ratio}. The error bars show the standard deviations from all 50 test simulations. The vertical lines denotes an observation ratio of 1.56\%, which is the same as in Figures \ref{fig:snapshots-enkf-srda} and \ref{fig:error-timeseries-enkf-srda}.}
\end{figure}

The accuracy of the ST-SRDA was comparable to that of the EnKF-HR with sparse observations. Figure \ref{fig:error-obs-ratio-enkf-srda} shows the dependence of the MAE ratio and MSSIM loss on the observation point ratio. The error bars represent the standard deviations based on all 50 test simulations. Both errors of the ST-SRDA (orange solid lines) are smaller than those of the EnKF-SR (green dashed lines), but larger than those of the EnKF-HR (blue dotted lines), as in Figure \ref{fig:error-timeseries-enkf-srda}. For all the models, the means and standard deviations of the test scores tended to increase as observations became sparser. In particular, the error bars of the EnKF-HR were so large that they overlapped with those of the ST-SRDA at an observation point ratio of 0.69\%. This implies that the accuracy of the ST-SRDA is relatively robust and becomes comparable to that of the EnKF-HR when observations are sparse.

The computation time of ST-SRDA was much shorter than that of EnKF-SR and EnKF-HR. Table \ref{table:wall-time} compares the average wall times of all the models. For reference, the table also lists the wall times of the underlying HR and LR fluid models. The wall time of the ST-SRDA was approximately 22\% and 1.3\% of that of the EnKF-SR and the EnKF-HR, respectively, which highlights the computational efficiency of ST-SRDA.

\begin{table}[t]
    \caption{\label{table:wall-time} Average wall times of all the models for the vorticity evolution over $0 \le t \le 24$. The EnKF-SR and ST-SRDA employ the low-resolution (LR) fluid model to calculate the time evolution, while the EnKF-HR employs the high-resolution (HR) fluid model. The number of ensemble members for the EnKF-SR and EnKF-HR was set to 100. All the calculations, including the NN inference, were performed using a single process of an Intel Xeon Gold 6242R CPU.}
    \centering
    \begin{tabular}{l r}
    \hline
    Model name & Wall time [s] \\
    \hline
    HR fluid model & 696.3 \\
    LR fluid model & 28.9 \\
    \hline 
    EnKF-HR & 4415.6 \\
    EnKF-SR & 262.5 \\
    ST-SRDA & 57.2 \\
    \hline
    \end{tabular}
\end{table}

\subsection{\label{subsec:obs-feature} Features for Observations Including Missing Values}

In general, NNs can learn data representations suitable for specific problems \cite<e.g.,>{Bengio+2013TPAMI, Zeiler+2014ECCV}. Here, it is interesting to examine the representations for the observation data because of missing values. We show that the spatial range with the influence of observed grid points can increase in deeper layers.

\begin{figure}[t]
    \includegraphics[width=\textwidth]{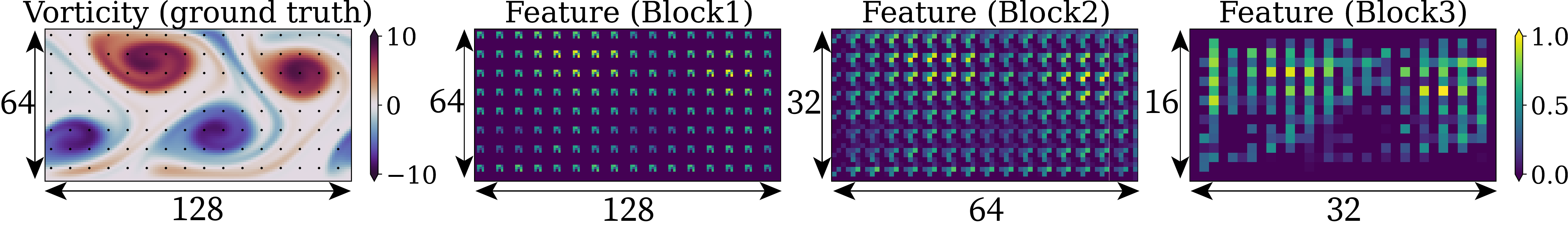}
    \caption{\label{fig:obs-feature} Features extracted from the observation data. The far-left panel shows a ground-truth snapshot, where the black dots represent the observed grid points. Since the average pooling is applied to the UHR data to generate the synthetic observations, the panel shows the ground-truth snapshot blurred by average pooling. In the observation data, the unobserved points were filled with the missing values, i.e., zeros (Section \ref{subsec:train-test-methods}). The other panels show the features extracted from the observations. These features are outputs from Block 1 through 3 in Figure \ref{fig:model-architecture}. The feature magnitudes are normalized to 1. Each number along the vertical or lateral axis denotes the data size.}
\end{figure}

Figure \ref{fig:obs-feature} shows features extracted from the observation data that appear to be explainable. These features are outputs from Block 1 through 3 in Figure \ref{fig:model-architecture} and have smaller spatial sizes in the deeper layers due to the U-Net architecture. In these feature fields, small background values may reflect missing values, while the larger values are interpreted as representations of each observed point. The wavenumber-2 structure in the ground truth becomes more obvious in deeper-level features (Figure \ref{fig:obs-feature}). This implies that the observed grid points may have a larger spatial influence in deeper layers.

\add{In the fourth panel, ``Feature (Block 3)'' is quite low near the center, which seems to coincide with the negative vortex in the first panel (ground truth). Moreover, this feature exhibits similar low values near unobserved points. These findings suggest that the NN has difficulty distinguishing between the missing values and minimum normalized vorticity, as both are represented by zeros} (Section \ref{subsec:neural-network}). \add{The differentiation can be achieved by introducing binary masks, which are frequently employed in image inpainting} \cite<e.g.,>{Liu+2018ECCV, Yu+2019ICCV}. \add{These masks assign $0$ to unobserved grid points and $1$ to observed points. Even if the missing and minimum observed values are the same, the binary mask enables the NN to distinguish between them. The image inpainting techniques may be beneficial for future studies.}

\subsection{\label{subsec:robustness-srda} Robustness of ST-SRDA Inferences}

Here, we demonstrate that the domain shift can reduce the accuracy of ST-SRDA and SR-mixup mitigates this degradation. The domain shift occurs due to the difference in the calculation process between the test and training phases. In the test phase, the NN inference and LR fluid simulation are performed alternately, where the LR fluid model is initialized by the NN inference (Section \ref{subsec:inference-4d-srda} and Figure \ref{fig:inference-train-method}a). In contrast, during the training, the LR fluid model is initialized using the HR simulation results, and feedback from the NN to the fluid model is not incorporated (Section \ref{subsec:training-4d-srda} and Figure \ref{fig:inference-train-method}b). We focused only on this difference in the calculation process by using the HR instead of the UHR ground truth. According to the training and test methods in Section \ref{subsec:train-test-methods}, the observations were generated from the HR simulations during training, whereas they were generated from the UHR results during testing. Thus, both the calculation process and method of generating observations differ from those in the training phase. This experimental setup is suitable for evaluating the ST-SRDA accuracy (Section \ref{subsec:accuracy-srda}), but not for investigating the influence of domain shift. To eliminate the difference in the observations, they are generated here from the HR ground truth. The evaluation metrics (Equations \ref{eq:mae-ratio} and \ref{eq:mssim}) are then computed in the HR space. We first examined that the difference in the calculation process leads to a domain shift, namely a statistical difference in the input of LR fluid-model states.

\begin{figure}[t]
    \includegraphics[width=\textwidth]{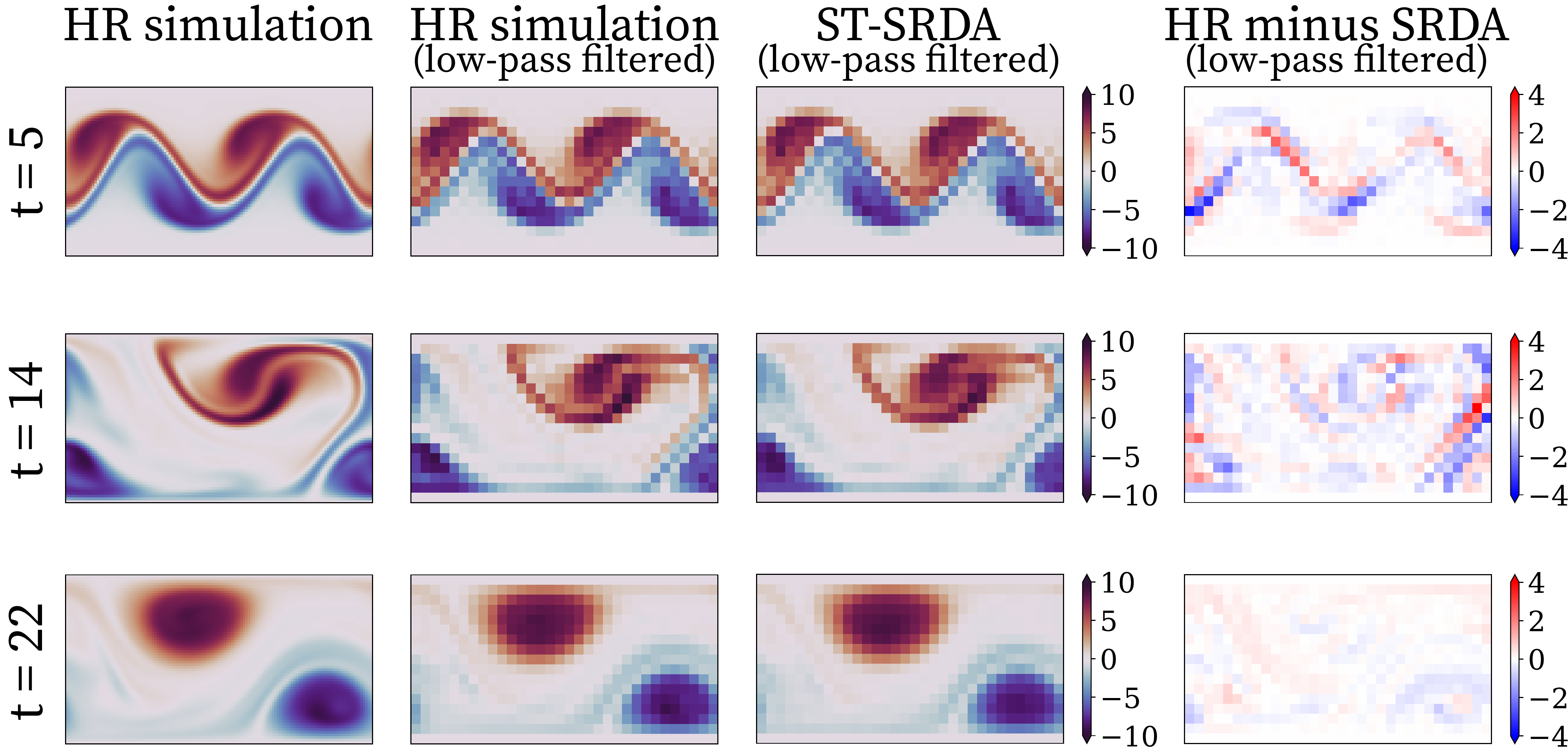}
    \caption{\label{fig:diff-lr-for-domain-shift} An example of the vorticity difference between the HR simulation and ST-SRDA inference. In the training phase, the LR fluid model was initialized by the low-pass filtered HR data, while in the testing phase, it was initialized by the filtered ST-SRDA inferences. The far-right panels show the differences between these two initial conditions. For reference, the unfiltered HR data are shown in the far-left panels.}
\end{figure}

Figure \ref{fig:diff-lr-for-domain-shift} shows an example of the vorticity difference between the HR simulation and ST-SRDA inference. In the training phase, the LR fluid model was initialized by the low-pass filtered HR simulation results, while in the testing phase, it was initialized by the filtered ST-SRDA inferences. The difference in the initial condition tended to be larger for the small-scale patterns associated with jet collapse ($t = 5$) and vortex merging ($t = 14$), whereas the difference was smaller in the final state with large-scale vortices ($t = 22$). We also confirmed that the probability density functions of vorticity values varied significantly in terms of the Kolmogorov-Smirnov test, where the p-value was less than $0.006$. These results indicate that the difference in the initial condition was significant, implying that the obtained LR time series was also statistically different between the training and test phases, i.e., a domain shift occurred.

We next investigated the influence of the domain shift. For comparison, the NN was trained without SR-mixup. Note that all results presented in Sections \ref{subsec:accuracy-srda} and \ref{subsec:obs-feature} employed SR-mixup. The NNs with and without SR-mixup were compared using the two evaluation methods illustrated in Figure \ref{fig:inference-train-method}. The first method (Figure \ref{fig:inference-train-method}a) is the same as the test method in Section \ref{subsec:accuracy-srda}, where the domain shift occurred. In contrast, in the second method (Figure \ref{fig:inference-train-method}b), the domain shift does not occur because it is the same as the method used to generate the training data.  In this method, there is no feedback from the NN inference to the LR simulation.

\begin{figure}[t]
    \includegraphics[width=\textwidth]{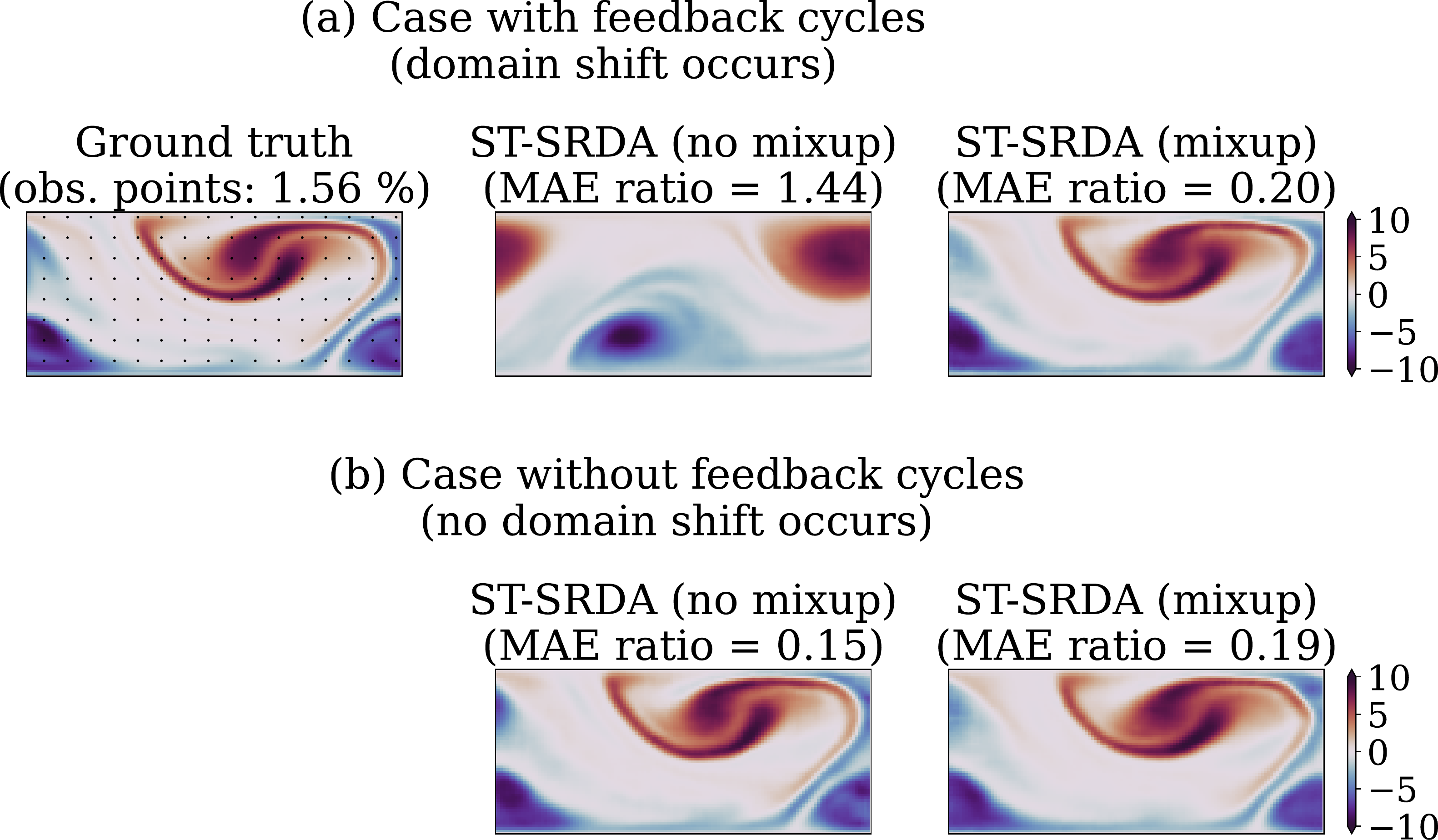}
    \caption{\label{fig:snapshots-srda-mixup} Vorticity snapshots from the ground truth and the inference of the ST-SRDA with and without SR-mixup; in (a), the feedback cycle was repeated for evaluation (Figure \ref{fig:inference-train-method}a), while in (b), there was no feedback cycle (Figure \ref{fig:inference-train-method}b). Only one snapshot of the ground truth is presented because it does not depend on the use of SR-mixup or the evaluation methods. The black dots in the ground-truth snapshot represent the observation points. The ratio (1.56\%) was calculated according to Equation \ref{eq:def-obs-ratio}.}
\end{figure}

The domain shift potentially reduces ST-SRDA performance, but SR-mixup can mitigate this effect. Figure \ref{fig:snapshots-srda-mixup} shows vorticity snapshots obtained using the two evaluation methods. Without SR-mixup, the vorticity was accurately inferred (MAE ratio $= 0.15$) when no domain shift occurred (Figure \ref{fig:snapshots-srda-mixup}b). However, when the feedback was incorporated, i.e., the domain shift occurred (Figure \ref{fig:snapshots-srda-mixup}a), the vortex shapes were completely different from those in the ground truth, resulting in a higher MAE ratio ($= 1.44$). In contrast, when SR-mixup was employed, the inferences were not strongly dependent on the presence or absence of the domain shift (Figures \ref{fig:snapshots-srda-mixup}a and \ref{fig:snapshots-srda-mixup}b, respectively). These results indicate that SR-mixup makes the inference robust to statistical differences in the input.

\begin{figure}[t]
    \includegraphics[width=\textwidth]{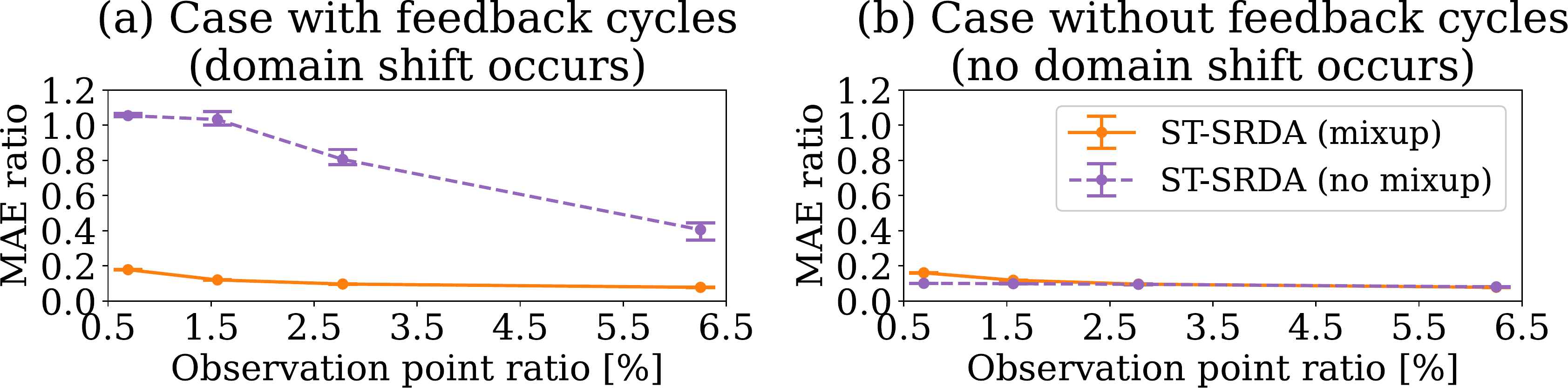}
    \caption{\label{fig:error-obs-ratio-srda-mixup} Dependence of the mean absolute error (MAE) ratio on the observation-point ratio with and without SR-mixup; in (a), the feedback cycle was repeated (Figure \ref{fig:inference-train-method}a), while in (b), there was no feedback cycle (Figure \ref{fig:inference-train-method}b). The MAE ratio is defined by Equation \ref{eq:mae-ratio}, and the values were averaged over time. The error bars show the maxima and minima of the MAE ratios from five experiments where the random seed for the initializer of neural-network weights was varied.}
\end{figure}

The effect of SR-mixup was further examined by varying the observation point ratio (Figure \ref{fig:error-obs-ratio-srda-mixup}). Without SR-mixup, the MAE ratio was substantially higher over the entire range of the observation point ratio when the domain shift occurred (dashed line in Figure \ref{fig:error-obs-ratio-srda-mixup}a). In contrast, using SR-mixup, the increase in the MAE ratio was suppressed, and the MAE ratio was almost independent of the domain shift (solid lines in Figures \ref{fig:error-obs-ratio-srda-mixup}a and \ref{fig:error-obs-ratio-srda-mixup}b). Interestingly, when the feedback was not incorporated (Figure \ref{fig:error-obs-ratio-srda-mixup}b), the MAE ratio was slightly higher with SR-mixup than without SR-mixup. During the training phase, SR-mixup added perturbations to the LR input (Equation \ref{eq:mixup-vorticity}), whereas in the evaluation for Figure \ref{fig:error-obs-ratio-srda-mixup}b, such perturbations were not added, and the LR initial condition was generated directly from the HR data. Thus, the training without SR-mixup resulted in smaller MAE ratios. However, this evaluation method is virtual because the LR fluid model is always initialized by the HR data; the inclusion of the feedback cycle is more realistic (Figure \ref{fig:error-obs-ratio-srda-mixup}a).

Furthermore, SR-mixup reduced the sensitivity of inference to the NN-weight initialization, as shown by the error bars in Figure \ref{fig:error-obs-ratio-srda-mixup}. In general, the NN inference can be influenced by randomly initialized weights \cite<e.g.,>{Bengio2012, Picard2021arXiv}. To assess this sensitivity, we trained the NN five times while changing the random seed for the weight initializer \cite{He+2015ICCV}. The error bars in the figure show the maxima and minima of the MAE ratios from these five experiments. When the domain shift did not occur, the error bars were quite small, regardless of SR-mixup (Figure \ref{fig:error-obs-ratio-srda-mixup}b). In contrast, when the domain shift occurred, the error bars were significantly larger without SR-mixup (dashed line in Figure \ref{fig:error-obs-ratio-srda-mixup}a). Therefore, SR-mixup suppressed the sensitivity and promoted the stability of the inference.

\begin{figure}[t]
    \includegraphics[width=\textwidth]{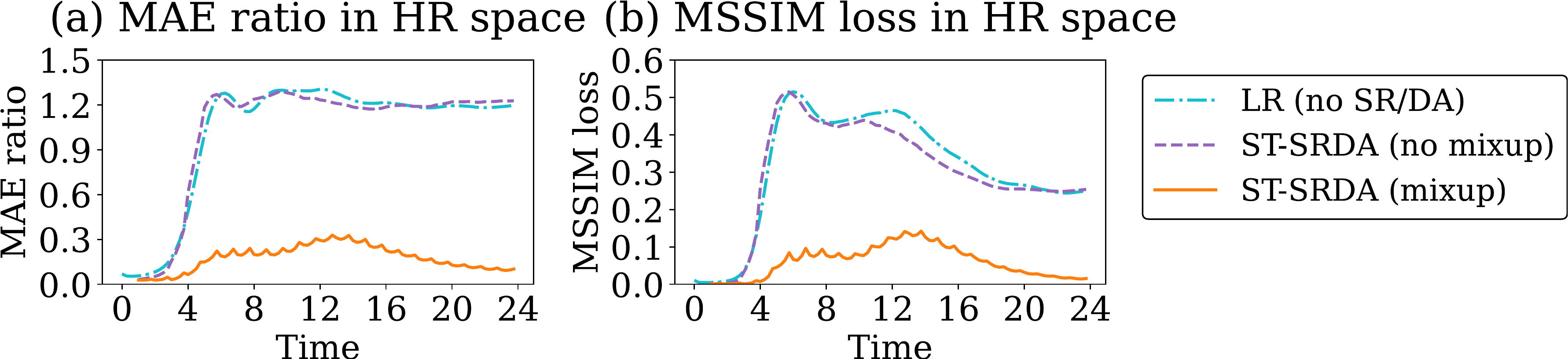}
    \caption{\label{fig:error-time-series-mixup} Time series of the test scores: (a) mean absolute error (MAE) ratio and (b) mean structural similarity index measure (MSSIM) loss. The MAE ratio and MSSIM loss are defined by Equation \ref{eq:mae-ratio} and \ref{eq:mssim}, respectively. The observation point ratio was set to 0.56\%. For comparison, the LR simulation was conducted without SR or DA, and its time series are labeled ``LR (no SR/DA).''}
\end{figure}

The MAE ratio without SR-mixup was saturated at an observation point ratio of 0.56\%, resulting in the small error bar (purple) in Figure \ref{fig:error-obs-ratio-srda-mixup}a. For comparison, we conducted the LR fluid simulation without DA or SR, which was expected to give an upper error bound for the ST-SRDA. Figure \ref{fig:error-time-series-mixup} compares the time series of the MAE ratio when the observation point ratio is 0.56\%. As expected, the MAE ratio of the ST-SRDA without SR-mixup was comparable to that of the LR simulation, both of which were much larger than that of the ST-SRDA with SR-mixup. In particular, the error grew rapidly in the initial period ($t \lesssim 4$). Similar error growth was observed at the other observation point ratios, i.e., with denser observations. During the initial period, the jet stream collapses while developing fine-scale vorticity filaments (Figure \ref{fig:vorticity-evolution}), which can enhance the domain shift (Figure \ref{fig:diff-lr-for-domain-shift}). This could explain the rapid error growth without SR-mixup found in Figure \ref{fig:error-time-series-mixup}.

Finally, we discuss three possible reasons why domain-generalization techniques have rarely been applied to fluid systems, including the atmosphere and ocean \cite{Brunton+2020ARFM, Duraisamy2021PRF, Kashinath+2021PTRS}. First, despite the reduced accuracy due to domain shift, the NN inference can remain more accurate than that of baselines, making domain-generalization methods less necessary. Second, loss functions that include physics-based terms, such as the residuals of the governing equations, can act as regularizers that improve generalizability \cite<e.g.,>{Jiang+2020ICHPC, Wang+2020NIPS, Bao+2022CUAI, Hammoud+2022JAMES}. Third, incorporating geometric symmetries as prior knowledge can enhance the generalization performance \cite<e.g.,>{Ling2016JFM, Wang2021ICLR, Chattopadhyay+2022GMDD, Yasuda+2023APLML}. Further research is needed to evaluate the validity of these factors. As we have demonstrated, domain shifts can occur when NNs are incorporated into physics-based models, thereby reducing accuracy. By utilizing domain-generalization techniques \cite<e.g.,>{Wang+2022IEEE, Zhou+2022IEEE}, even in the three cases described here, it may be possible to further improve the robustness of NNs and extend their applicability to different systems.

\section{Conclusions} \label{sec:conclusions}  

This study proposes a four-dimensional super-resolution data assimilation (4D-SRDA) scheme. This scheme employs a physics-based model to calculate the time evolution of a system using low-resolution (LR) simulations. The resulting time series of LR model states, along with observations, are fed into a trained neural network (NN). This NN simultaneously performs data assimilation (DA) and spatio-temporal super-resolution (SR), generating flow fields with high spatio-temporal resolution. A snapshot of these high-resolution (HR) flow fields is considered the analysis and projected onto the LR grid points via algebraic operations, such as linear interpolation. Using this LR analysis, an LR simulation is conducted for the next assimilation cycle. 4D-SRDA does not require ensemble evolution or iterative calculations, making it computationally efficient.

In the 4D-SRDA system, the NN is trained via supervised or unsupervised learning, in which a domain shift, namely a statistical difference between the training and test data, can affect the performance of the system \cite<e.g.,>{MorenoTorres+2012PR}. In general, when physics-based simulations and NN inferences are performed alternately and are mutually dependent in the test phase, a domain shift will occur if the training data are generated in advance solely by running the physics-based model (i.e., offline training).

To prevent the degradation in accuracy caused by the domain shift, we developed a data augmentation method for domain generalization, SR-mixup, based on C-Mixup \cite{Yao+2022NIPS}. In SR-mixup, two similar inputs are randomly sampled and their linear combination is used as input to improve robustness to domain shift. Unlike C-Mixup, the linear combination is not made for the output, to avoid producing blurred HR images in SR problems. SR-mixup is applicable to supervised and unsupervised problems because it does not assume paired inputs and outputs.

We validated the 4D-SRDA and SR-mixup using an idealized barotropic ocean jet \cite{David+2017OM} with supervised learning. Since this jet was modeled as a two-dimensional flow, 4D-SRDA was referred to as spatio-temporal SRDA (ST-SRDA). In the case of supervised learning, the ST-SRDA is regarded as a computationally efficient surrogate scheme for generating an HR reanalysis dataset. For comparison, we applied an ensemble Kalman filter (EnKF) to the LR or HR fluid model. The former is a baseline model for the ST-SRDA, while the latter is considered to give a lower error bound for the ST-SRDA because the underlying HR fluid model generates the HR training data. The ST-SRDA always tended to be more accurate than the baseline, i.e., the EnKF at the LR. When the observations were sparse, the error of the ST-SRDA became comparable to that of the EnKF at the HR (i.e., close to a lower error bound). Furthermore, the wall time of ST-SRDA was much shorter than that of the EnKFs. These results suggest that ST-SRDA is effective for generating HR reanalysis at low computational cost. We also confirmed that the error became substantially larger without SR-mixup, indicating that the combination of SR-mixup and ST-SRDA is useful in achieving robust forecast cycles.

To improve the applicability of this scheme to real-world contexts, a training method using small patches should be developed. In this study, the NN was trained based on the whole two-dimensional flow fields. The data size of three-dimensional flows is larger, which requires the division of a snapshot into small patches. Training methods with small patches have been developed in terms of DA \cite{Ouala+2018RS, Yang+2021JCP} and may be effective in applying 4D-SRDA to more realistic three-dimensional data.

Another direction for future work is to achieve 4D-SRDA with unsupervised learning. The proof-of-concept presented here employed supervised learning, assuming that sufficiently accurate atmospheric or oceanic data were available in the training phase. In practice, such data can be obtained from HR reanalysis datasets. Unsupervised settings will make 4D-SRDA applicable to a broader range of conditions, in which accurate data may not be available. Furthermore, since 4D-SRDA does not require ensembles, it is difficult to estimate the uncertainty of the inferences. This may be improved by utilizing unsupervised-learning frameworks \cite<e.g.,>{Krishnan+2015arXiv, Fraccaro+2017NIPS}. Although the numerical experiments in this paper are idealized, the super-resolution and domain-generalization techniques presented herein provide a useful basis for future developments in the integration of data-driven and physics-based models.

\appendix

\section{\label{sec:hyper-param-nn}Hyperparameters of the Neural Network}

We list here the hyperparameters of the NN (Figure \ref{fig:model-architecture}), which were determined through a grid search. Further details about the implementation are available at the Zenodo repository (see Open Research). Tables \ref{table:hyper-param-encoder} and \ref{table:hyper-param-decoder} list the layer-dependent hyperparameters in the encoder and decoder, respectively. The kernel size of all convolution layers is $3$. The slope of all leaky ReLU layers is $-0.01$ \cite{Maas+2013ProcICML}. The upscale factor of all pixel shuffle layers is $2$ \cite{Shi+2016CVPR}. The hyperparameters in the transformer encoding blocks are set as follows \cite{Vaswani+2017NIPS}: the number of input features is $2048$, the number of heads in the multi-head attention is $16$, and the dimension of the feed-forward network is $2048$.

\begin{table}[t]
    \caption{\label{table:hyper-param-encoder} Hyperparameters in the encoder. The block and layer names are defined in Figure \ref{fig:model-architecture}.}
    \centering
    \begin{tabular}{l l r r r}
    \hline
    Block name & Layer name & Input channels & Output channels & Stride \\
    \hline
    ---                           & Conv2D-1 &   1 &  64 & 1 \\
    \hline
    \multirow{2}{*}{Downsample-1} & Conv2D-2 &  64 & 128 & 2 \\
                                  & Conv2D-3 & 128 & 128 & 1 \\
    \hline
    \multirow{2}{*}{Downsample-2} & Conv2D-2 & 128 & 256 & 2 \\
                                  & Conv2D-3 & 256 & 256 & 1 \\
    \hline
    \multirow{2}{*}{Downsample-3} & Conv2D-2 & 256 & 128 & 2 \\
                                  & Conv2D-3 & 128 & 128 & 1 \\
    \hline
    \multirow{2}{*}{Downsample-4} & Conv2D-2 & 128 &  64 & 2 \\
                                  & Conv2D-3 &  64 &  64 & 1 \\
    \hline
    \end{tabular}
\end{table}

\begin{table}[t]
    \caption{\label{table:hyper-param-decoder} Hyperparameters in the decoder. The block and layer names are defined in Figure \ref{fig:model-architecture}.}
    \centering
    \begin{tabular}{l l r r}
    \hline
    Block name & Layer name & Input channels & Output channels \\
    \hline
    \multirow{2}{*}{Upsample-1} & Conv2D-4 &  64 & 256 \\
                                & Conv2D-5 &  64 & 128 \\
    \hline
    \multirow{2}{*}{Upsample-2} & Conv2D-4 & 128 & 512 \\
                                & Conv2D-5 & 128 & 256 \\
    \hline
    \multirow{2}{*}{Upsample-3} & Conv2D-4 & 256 &1024 \\
                                & Conv2D-5 & 256 & 128 \\
    \hline
    \multirow{2}{*}{Upsample-4} & Conv2D-4 & 128 & 512 \\
                                & Conv2D-5 & 128 &  64 \\
    \hline
    \multirow{2}{*}{---}        & Conv2D-6 & 192 &  64 \\
                                & Conv2D-7 &  64 &   1 \\
    \hline
    \end{tabular}
\end{table}

\section*{Open Research}

The source code used for generating all the datasets and experiments is preserved at \url{https://doi.org/10.5281/zenodo.7946486} and developed openly at \url{https://github.com/YukiYasuda2718/4d-srda_sr-mixup}. The detailed information can be found in \citeA{Yasuda+2023ZenodoSrdaMixup}.

\acknowledgments
\add{We greatly appreciate the three anonymous reviewers, the associate editor, and the editor for their careful review and constructive feedback on the manuscript. }This work was supported by the JSPS KAKENHI (Grant Number 20H05751). This work used computational resources of the TSUBAME3.0 supercomputer provided by the Tokyo Institute of Technology through the HPCI System Research Project (Project ID: hp220102). This paper is based on results obtained from a project, JPNP22002, commissioned by the New Energy and Industrial Technology Development Organization (NEDO).


%
%



\bibliography{reference}

%
%
%
%
%

\end{document}